\documentclass[a4paper,fleqn,usenatbib]{mnras}

\usepackage{newtxtext,newtxmath}

\usepackage[T1]{fontenc}
\usepackage{lmodern}

\usepackage{graphicx}	
\usepackage{amsmath}	
\usepackage{amssymb}	
\usepackage{float}

\title[NIR photometry of faint transients]{Near-infrared counterparts of three transient very faint neutron star X-ray binaries}

\author[A. W. Shaw et al.]{
A. W. Shaw$^{1}$\thanks{E-mail: aarran@ualberta.ca}
C. O. Heinke,$^{1}$
N. Degenaar,$^{2}$
R. Wijnands,$^{2}$
R. Kaur,$^{3}$
L. M. Forestell$^{1,4,5}$
\\
$^{1}$Department of Physics, University of Alberta, CCIS 4-181, Edmonton, AB T6G 2E1, Canada\\
$^{2}$Anton Pannekoek Institute for Astronomy, University of Amsterdam, Postbus 94249, Science Park 1098 XH, Amsterdam, Netherlands\\
$^{3}$Physics Department, Suffolk University, 41 Temple Street, Boston, Massachusetts, 02114, USA\\
$^{4}$Department of Physics and Astronomy, University of British Columbia, Vancouver, BC V6T 1Z1, Canada\\
$^{5}$TRIUMF, 4004 Wesbrook Mall, Vancouver, BC V6T 2A3, Canada\\
}
\date{Accepted XXX. Received YYY; in original form ZZZ}

\pubyear{2015}

\begin{document}
\label{firstpage}
\pagerange{\pageref{firstpage}--\pageref{lastpage}}
\maketitle

\begin{abstract}
We present near-infrared (NIR) imaging observations of three transient neutron star X-ray binaries, SAX J1753.5--2349, SAX J1806.5--2215 and AX J1754.2--2754. All three sources are members of the class of `very faint' X-ray transients which exhibit X-ray luminosities $L_X\lesssim10^{36}$ erg s$^{-1}$. The nature of this class of sources is still poorly understood. We detect NIR counterparts for all three systems and perform multi-band photometry for both SAX J1753.5--2349 and SAX J1806.5--2215, including narrow-band Br$_{\gamma}$ photometry for SAX J1806.5--2215. We find that SAX J1753.5--2349 is significantly redder than the field population, indicating that there may be absorption intrinsic to the system, or perhaps a jet is contributing to the infrared emission. SAX J1806.5--2215 appears to exhibit absorption in Br$_{\gamma}$, providing evidence for hydrogen in the system.  Our observations of AX J1754.2--2754 represent the first detection of a NIR counterpart for this system. We find that none of the measured magnitudes are consistent with the expected quiescent magnitudes of these systems. Assuming that the infrared radiation is dominated by either the disc or the companion star, the observed magnitudes argue against an ultracompact nature for all three systems.
\end{abstract}




\section{Introduction}
\label{sec:Intro}
Low-mass X-ray binaries (LMXBs) are binary systems in which a compact object, either a black hole (BH) or neutron star (NS), accretes matter from a low-mass star. Many LMXBs are discovered when they undergo transient outbursts, where the X-ray luminosity increases by a factor $>1000$, accompanied by a large increase in optical luminosity \citep[$\Delta V\sim7$; e.g.,][]{Kuulkers-1998} for a short period (typically weeks to months) before decaying to quiescence. \\
\indent Whilst many transient LMXBs exhibit high luminosities during outburst ($L_X\sim10^{37-39}$ erg s$^{-1}$), during the last decade, a population of faint X-ray transients have emerged. Members of this class of Very Faint X-ray Transients (VFXTs) exhibit peak X-ray luminosities in the range $10^{34-36}$ erg s$^{-1}$ \citep[e.g. ][]{Muno-2005a,Wijnands-2006,Degenaar-2009,Degenaar-2010b}. Due to the frequent monitoring of the Galactic Centre, a large number of VFXTs have been found to lie close to this region.\\
\indent The current disc instability model (DIM) which well describes the general behaviour of the outbursts in typical LMXB transients \citep{Lasota-2001,Coriat-2012} cannot immediately explain the low peak X-luminosities of VFXTs \citep{Heinke-2015,Hameury-2016}. The luminosities of these sources time-averaged accretion rate in the range $10^{-13}-10^{-10}M_{\odot}{\rm yr}^{-1}$ \citep{King-2006,Degenaar-2009}, which can be difficult to explain in the context of binary evolution models. A number of models have emerged in an effort to characterize the observed low accretion rates and faint X-ray luminosities. It is possible that some VFXTs are wind accreting systems, similar to high-mass X-ray binary (HMXB) systems, except that the compact object is accreting from the weak stellar wind of a low-mass companion \citep{Pfahl-2002,Maccarone-2013}. An alternative model is that the compact object has a small accretion disc, indicative of an ultracompact system with a short orbital period \citep[$P_{{\rm orb}}\lesssim2$hrs; e.g.][]{int-Zand-2005,int-Zand-2007,Hameury-2016} that can only accommodate a (partly) degenerate donor such as a brown or white dwarf \citep{King-2006,Heinke-2015}. It has also been proposed that the magnetic field of a NS can inhibit accretion, resulting in the observed very faint X-ray luminosities \citep{Heinke-2009,DAngelo-2012,Heinke-2015}. Some VFXTs have been shown to have a large orbital inclination, meaning observers only see the scattered light of an intrinsically bright source \citep{Muno-2005b,Corral-Santana-2013}, but this scenario cannot be applied to most systems \citep{Wijnands-2006}.\\
\indent The nature of the donors in VFXTs can be best understood through optical/near infrared (NIR) follow-up of sources discovered in X-ray surveys. However, as most of the sources have been found so far to be close to the Galactic Centre, optical/NIR photometry is inhibited by high extinction and extremely crowded fields \citep{Mauerhan-2009}. Hence, optical/NIR results have not been reported for most VFXTs as it becomes difficult to identify the correct counterpart. Despite this, there are a small number of observations of VFXTs in the optical/NIR regime. \citet{Degenaar-2010a} identified the optical counterpart of the bursting NS binary 1RXH J173523.7--354013, revealing H$_{\alpha}$ emission in the optical spectrum and effectively ruling out an ultracompact nature \citep[though it must be noted that recent efforts to model ultracompact evolution have provided evidence that some ultracompact systems can potentially exhibit hydrogen in their spectra;][]{Sengar-2017}. Through optical observations, M15 X-3 has been shown to contain a main sequence star in a $\sim4$ hr orbit with a NS \citep{Heinke-2009,Arnason-2015}. In addition, XTE J1719--291, Swift J1357.2--0933, SAX J1806.5--2215 and CXOGC J174540.0--290031 have all been suggested to contain main-sequence companions from optical/NIR observations \citep{Muno-2005b,Greiner-2008,Corral-Santana-2013,Kaur-2017}.\\
\indent Based on their optical/NIR properties, all of the above sources appear to be typical LMXBs, not ultracompact systems. However, owing to the small sample size, we cannot yet rule out any of the above models. We need to investigate the companion stars of many more VFXTs in order to better understand the different possible accretion regimes in these systems. Due to the typically large absorption columns in the Galactic plane, searching for counterparts at NIR wavelengths is much more effective than at optical wavelengths. In this work we present NIR photometry of three VFXTs in order to place constraints on their counterparts and investigate the nature of accretion in these systems.

\subsection{SAX J1753.5--2349}
SAX J1753.5--2349 (hereafter SAX1753) was discovered with the Wide Field Camera (WFC) instrument on board the \emph{BeppoSAX} X-ray observatory during a single type-I burst on August 24 1996 \citep{int-Zand-1999}. Initially a `burst-only' source, persistent emission (either in outburst or quiescence) was not detected for SAX1753 until the source was once more detected in outburst in 2008 by the Proportional Counter Array on board the \emph{Rossi X-ray Timing Explorer} (\emph{RXTE}/PCA) and the \emph{Swift}/Burst Alert Telescope (BAT) \citep{Markwardt-2008}, as well as the \emph{INTErnational Gamma-Ray Astrophysics Laboratory} (\emph{INTEGRAL}) imager IBIS \citep{CadolleBel-2008}. A further X-ray burst and associated transient outburst was observed by \emph{INTEGRAL} and \emph{RXTE}/PCA in 2010 \citep{Chenevez-2010} and a NIR counterpart with $K_s =15.63\pm0.01$ was identified \citep{Torres-2010}. \\
\indent Modelling of the broadband X-ray spectrum of SAX1753 from the 2008 outburst \citep[which lasted $>5$ months;][]{DelSanto-2009} revealed a spectrum consistent with the Compton up-scattering of soft seed photons by a hot optically thin electron plasma with an inferred temperature $\gtrsim24$ keV \citep{DelSanto-2010}. The low luminosity of SAX1753 (peaking at $L_X\sim10^{36}$erg s$^{-1}$) suggests that the system is very compact\citep{DelSanto-2010}, though its orbital period has not yet been measured and, as discussed in Section \ref{sec:Intro}, there are number of alternative possible origins for the low X-ray luminosity. The compact object is known to be a neutron star from observations of thermonuclear bursts \citep{Chakrabarty-2010}. \\

\subsection{SAX J1806.5--2215}
SAX J1806.5--2215 (hereafter SAX1806) was discovered by \emph{BeppoSAX}/WFC through the detection of four type-I X-ray bursts observed between August 1996 and October 1997 \citep{int-Zand-1999,Cornelisse-2002b}. Similar to SAX1753, SAX1806 was initially classified as a `burst-only' source until faint persistent emission was revealed by the \emph{RXTE}/All Sky Monitor (ASM) coinciding with the same period as the occurrence of the X-ray bursts \citep{Cornelisse-2002b}. \\
\indent The source remained quiescent for 12 yrs, with upper limits on the X-ray luminosity during this time estimated to be $(0.5-4) \times 10^{33}$ erg s$^{-1}$ \citep{Campana-2009,Degenaar-2011}. A new outburst was detected by \emph{RXTE} in February 2011 \citep{Altamirano-2011}. \emph{Swift} observations revealed an X-ray spectrum consistent with a power law with a hard photon index $\Gamma\sim1.7-2$ \citep{Degenaar-2011,Kaur-2012b,DelSanto-2012}. Monitoring with \emph{Swift}/BAT indicates that the source is still exhibiting low-level activity and hence has likely remained in outburst for $\gtrsim6$ years.\footnote{https://swift.gsfc.nasa.gov/results/transients/weak/SAXJ1806.5-2215/} During the X-ray outburst that began in 2011, a NIR counterpart was discovered with $K=17.25\pm0.03$ \citep{Kaur-2017}.\\

\subsection{AX J1754.2--2754}
AX J1754.2--2754 (hereafter AX1754) was discovered in 1999 by the \emph{Advanced Satellite for Cosmology and Astrophysics} (\emph{ASCA}) during a survey of the Galactic Centre region \citep{Sakano-2002}. A type-I X-ray burst in April 2005 revealed the compact object as a NS \citep{Chelovekov-2007a}. The source has been detected at a luminosity of $L_x\sim10^{35}$ erg s$^{-1}$ every time it has been observed \citep[see e.g.][]{Jonker-2008,Degenaar-2012,Maccarone-2012}, aside from one brief ($\lesssim11$ months) period of quiescence \citep[$L_X\lesssim5\times10^{32}$ erg s$^{-1}$;][]{Bassa-2008}. A long observing campaign of AX1754 with \emph{Swift} revealed a very soft X-ray spectrum \citep[$\Gamma=2.5$;][]{Armas-Padilla-2013}.\\
\indent Unlike those of SAX1753 and SAX1806, the X-ray bursts displayed by AX1754 are long \citep[up to 15 minutes in length;][]{Chelovekov-2007b,Chenevez-2017}. Intermediate-to-long duration bursts are thought to be due to the ignition of a thick layer of helium on the surface of the NS which builds up due to the low mass accretion rate \citep{Peng-2007,Cooper-2007}. This, coupled with the non-detection of the source at optical/NIR wavelengths \citep{Bassa-2008,Zolotukhin-2015} suggest that AX1754 is an ultracompact X-ray binary, though no orbital period has been measured.\\

\section{Observations and Data Reduction}
\subsection{\emph{Swift}}
We used all available \emph{Swift}/X-Ray Telescope \citep[XRT;][]{Burrows-2005} observations of each source to construct long-term X-ray light curves in order to put the NIR observations into context. The light curves were created with the online \emph{Swift}/XRT User Objects tool \citep{Evans-2009} and grouped into 1 day bins. In the event of a non-detection, a $3\sigma$ upper limit on the XRT count rate is calculated by the light curve generator. We also used the data from the \emph{Swift}/BAT Hard X-ray Transient Monitor \citep{Krimm-2013} to construct long term X-ray light curves in the 15-50 keV range. Data are available online and do not require any reduction.\footnote{https://swift.gsfc.nasa.gov/results/transients/}\\

\subsection{Near Infrared}
All three sources were observed with the Near InfraRed Imager and spectrograph \citep[NIRI;][]{Hodapp-2003} on the 8.1m \emph{Gemini} North telescope at Mauna Kea, Hawaii with the f/6 camera in imaging mode. The f/6 camera has a plate scale of $0.117''$ pixel$^{-1}$, providing a field of view of $120''\times120''$. SAX1753 was observed on the night of 2012 July 11. We obtained 34 exposures of 53s in $H$ and 45 exposures of 56s in $K_s$. To account for the changing sky background at NIR wavelengths, a dithering pattern was applied in each filter, with each co-added exposure consisting of 17, 3s exposures in $H$ and 20, 2.7s exposures in $K_s$. \\
\indent SAX1806 was observed on 2012 May 2. The source was observed in $H$ and $K_s$ broad-band filters as well as the Br$_{\gamma}$ narrow-band filter. We obtained 10 exposures of 39s in both $H$ and Br$_\gamma$, with each co-add comprising of three, 13s dithered exposures. 10 exposures of 12s were obtained in $K_s$, with each co-add comprising of three, 4s dithered exposures. \\
\indent AX1754 was observed with \emph{Gemini}/NIRI on 2012 May 21. The field was observed in $H$ and $K_s$ broad-band filters as well as the Br$_{\gamma}$ narrow-band filter. However, as detailed in section \ref{AX1754_res}, we were only able to detect the counterpart in the $K_s$-band. We therefore choose to concentrate here on the $K_s$-band observations. We obtained 10 exposures of 12s, each comprising of three, 4s dithered exposures. \\
\indent Data reduction for all NIRI images is performed using the \textsc{iraf} \citep{IRAF} \emph{Gemini} package and NIRI-specific \textsc{python} routines. To remove artifacts superimposed by the IR detector controller we utilized the \textsc{cleanir} script\footnote{http://staff.gemini.edu/$\sim$astephens/niri/cleanir/cleanir.py} before correcting for non-linearity in the detector with \textsc{nirlin}.\footnote{http://staff.gemini.edu/$\sim$astephens/niri/nirlin/} As there were no extended objects in the fields of view of each target, sky frames were created from the science images. For each target, a normalized flat field was created with the task \textsc{niflat} and bad pixels identified using short dark frames. Flat-fielding and sky subtraction was performed using \textsc{nireduce} and the final images were created with \textsc{imcoadd}. From the final reduced images we obtained a signal-to-noise ratio, ${\rm S/N} > 10$ for a 17th magnitude star, which agrees with the estimates made by the NIRI Integration Time Calculator (ITC).\footnote{http://www.gemini.edu/sciops/instruments/integration-time-calculators/niri-itc}\\
\indent To determine the astrometry, we first used SExtractor \citep{Bertin-1996} to create source catalogs for each target field. The astrometric solution in each band band was obtained with SCAMP \citep{Bertin-2006}, using the Two Micron All Sky Survey \citep[2MASS;][]{Skrutskie-2006} catalog as a reference, then the projection was applied to the co-added science images using SWarp \citep{Bertin-2002}. The solution delivered an RMS error $\sim0.1''$ in each coordinate, indicative of the positional uncertainty.\\
\indent To measure instrumental magnitudes we used the \textsc{iraf daophot} routines developed for crowded-field photometry \citep{Stetson-1987}. We determined the empirical point spread function (PSF) for each target frame using relatively isolated field stars. We identified 10 suitable PSF stars in each band for the images of SAX1753 and AX1754, and, due to extreme crowding and a large number of saturated stars, 4 suitable PSF stars in each band for the SAX1806 images. The PSF was then fit to all the stars in the frame with the task \textsc{allstar} to determine instrumental magnitudes. The observations were calibrated using 2MASS stars in the field. We used a total of 10 calibration stars for SAX1753, 4 for SAX1806 (due to, again, crowding and saturated stars, as well as a difference in pointing between $H$ and $K_s$ observations such that only 4 suitable calibration stars were present in both bands) and 6 for AX1754. For SAX1753 and SAX1806 we calibrated the frames using transformation equations of the form:
\begin{equation}
h = H + c_1 + c_2X + c_3(H - K_s)
\end{equation}
\begin{equation}
k_s =  K_s + c_4 + c_5X + c_6(H - K_s)
\end{equation}
\noindent where $h$ ($k_s$) is the instrumental magnitude of the calibration star in the $H$ ($K_s$)-band, $H$ ($K_s$) is its known magnitude, $c_1$ -- $c_6$ are constants (representing an additive term, an extinction term and a colour term in each band), $X$ is the airmass and $H-K_s$ is the known NIR colour of the calibration star. \\
\indent For AX1754, we were only able to detect a counterpart in the $K_s$-band. We therefore calibrated its magnitude by determining the zero-point for our observations using the 6 well-detected, isolated 2MASS stars in the field.

\section{Results}
\subsection{SAX J1753.5-2349}
\label{res:1753}

\begin{figure*}
	\centering
	\includegraphics[width=0.75\textwidth]{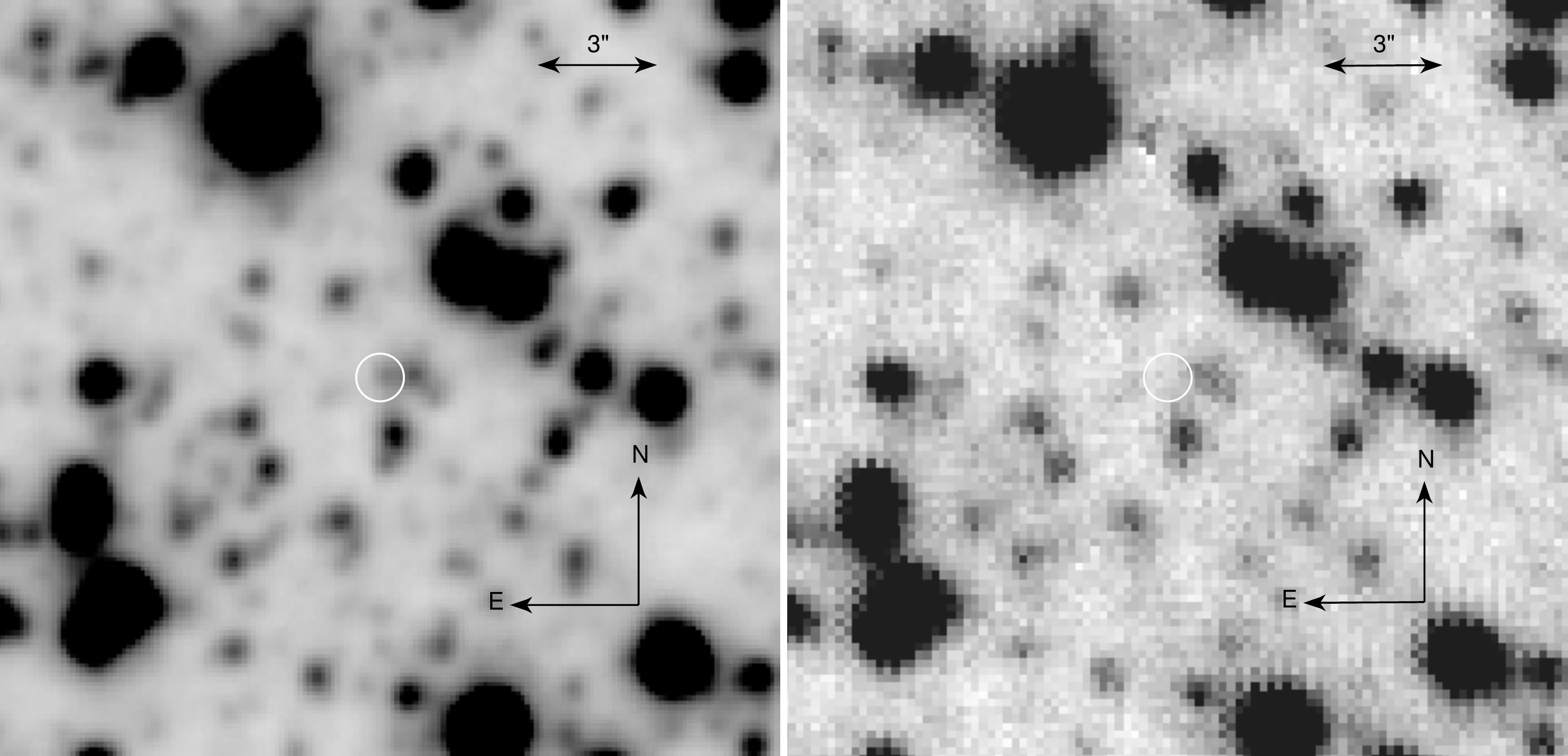}
	\caption{\emph{Left:} $K_s$-band NIRI image of SAX1753. \emph{Right:} $K$-band image of the same field, observed on 2007 May 4 as part of the UKIDSS. In both panels, the white circle represents the 0.6$''$ error circle of the \emph{Chandra} X-ray position of the source \citep{Chakrabarty-2010}.}
	\label{SAX1753_image}
\end{figure*}

\begin{figure}
	\centering
	\includegraphics[width=0.48\textwidth]{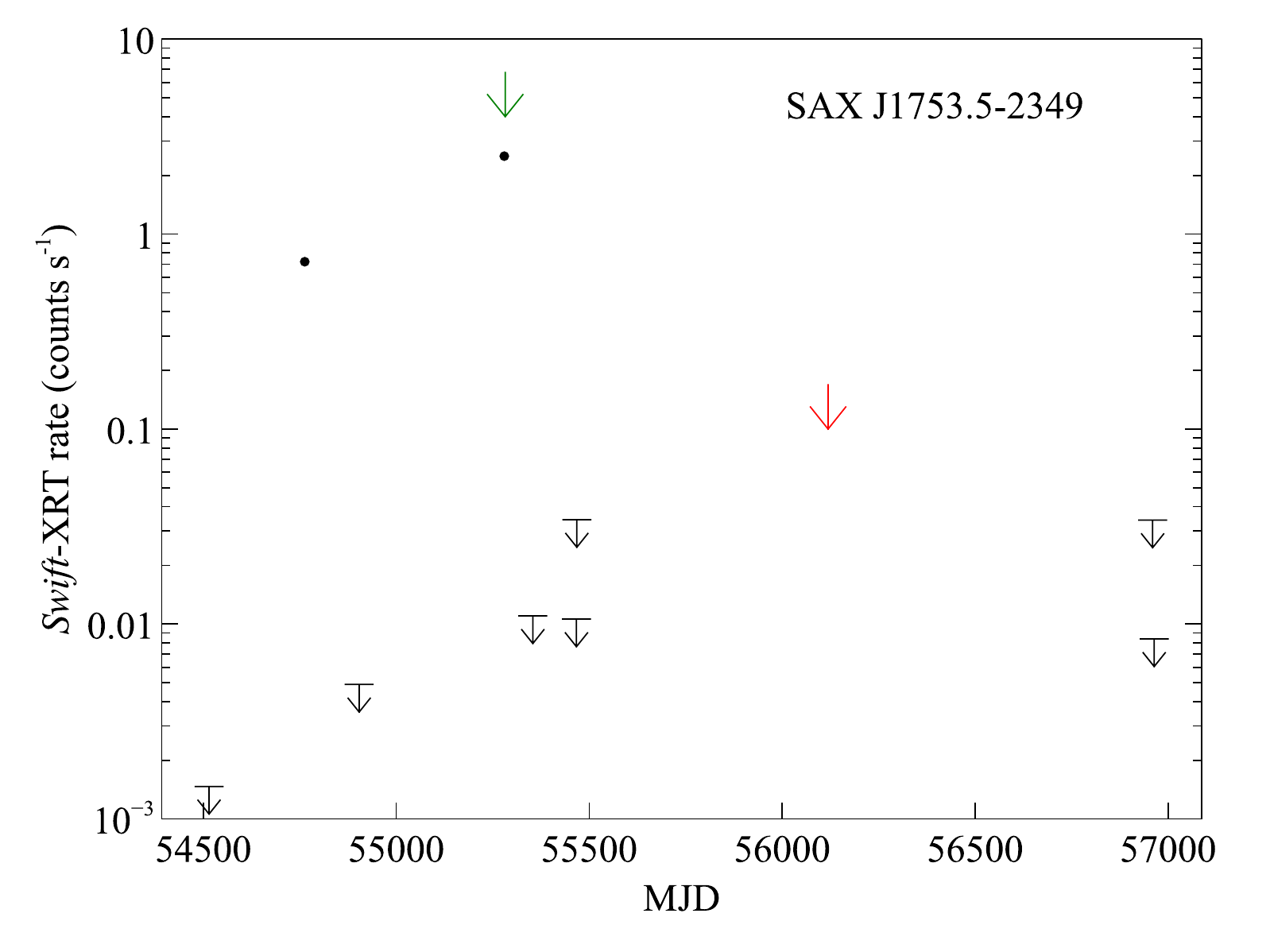}
	\caption{Long term \emph{Swift}/XRT light curve of SAX1753, the green arrow represents the time of the observation presented by \citet{Torres-2010}, the red arrow indicates the time of the \emph{Gemini}/NIRI observation. Black, capped arrows represent $3\sigma$ upper limits on the XRT count rate.}
	\label{SAX1753_Swift}
\end{figure}

\begin{figure}
	\centering
	\includegraphics[width=0.5\textwidth]{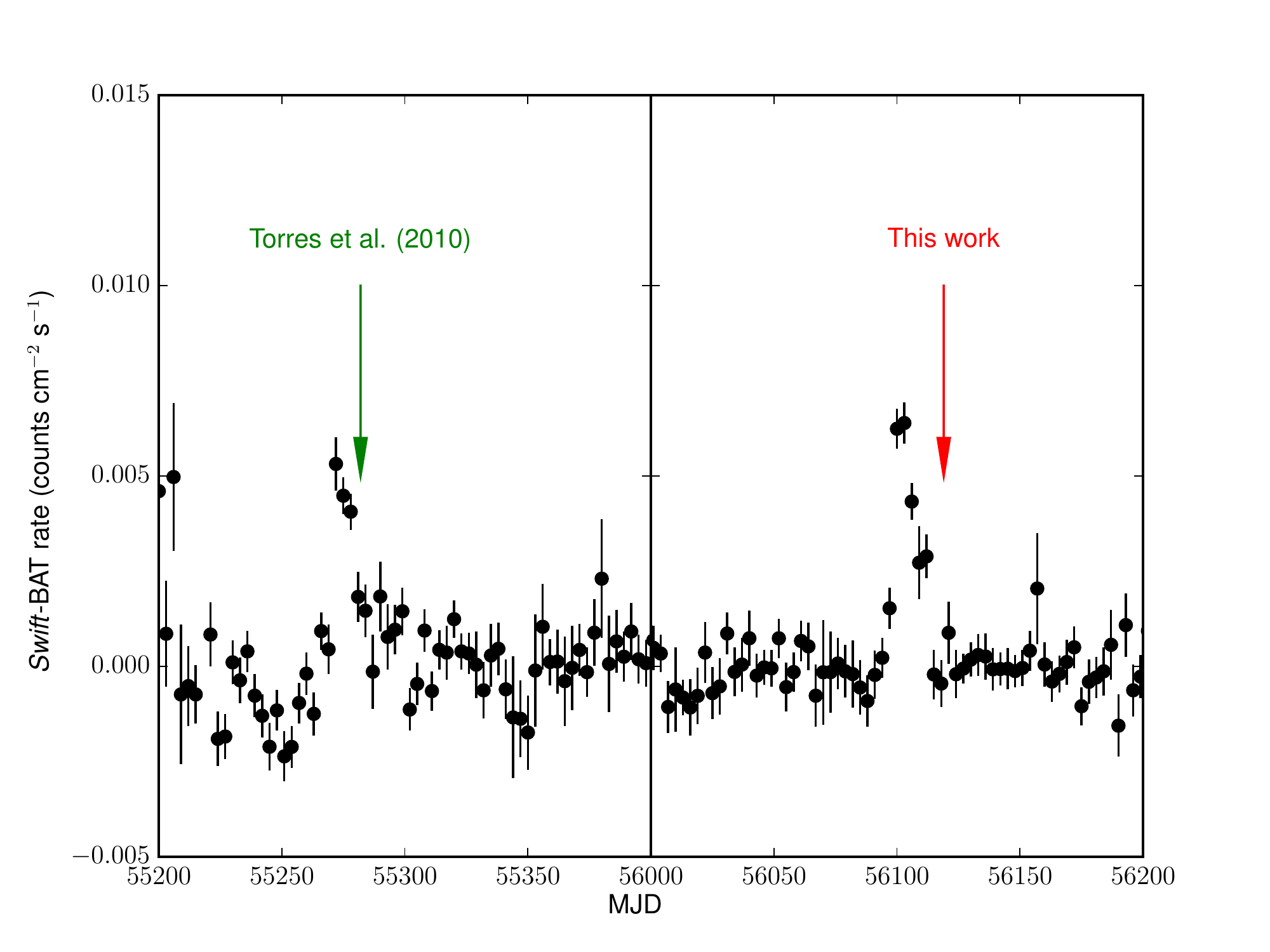}
	\caption{Portions of the long term \emph{Swift}/BAT light curve of SAX1753. The green arrow in the left panel indicates the time of the NIR follow-up observation by \citet{Torres-2010}, the red arrow in the right panel indicates the time of the \emph{Gemini}/NIRI observation presented in this work.}
	\label{SAX1753_BAT}
\end{figure}

Fig. \ref{SAX1753_image} shows the $K_s$-band NIRI image of the field of SAX1753, as well as an archival image observed with the \emph{United Kingdom Infrared Telescope} (\emph{UKIRT}) as part of the \emph{UKIRT} Infrared Deep Sky Survey (UKIDSS). We detect the source in the NIRI image at a position of RA, Dec = $17^{\mathrm{h}}53^{\mathrm{m}}31^{\mathrm{s}}.87$, $-23^{\circ}49'14''.83$, consistent with the position of the source determined during outburst using a \emph{Chandra} observation \citep{Chakrabarty-2010} and from the NIR counterpart \citep{Torres-2010}. We measure magnitudes of the NIR counterpart of $H=18.58\pm0.03$ and $K_s=17.44\pm0.02$, which is $\sim2$ magnitudes fainter than during the 2010 outburst \citep{Torres-2010}. Though this may indicate that the source was quiescent at the time, it is important to note that these magnitudes would have been measurable in the UKIDSS data, which has $5\sigma$ limits of $K=18.05$, $H=19.00$ \citep{Lucas-2008}. In addition there are sources in the UKIDSS field fainter than the measured NIRI magnitudes of SAX1753 \citep{Lucas-2008}, yet SAX1753 is not detectable by UKIDSS. This suggests that SAX1753 was not in true quiescence during the time of our observations, but was still active at a low level. The X-ray light curve presented in Fig. \ref{SAX1753_Swift} suggests that the source had reached quiescence at X-ray energies. However, there were no pointed X-ray observations at the time of the NIRI observations. We therefore investigated the long-term \emph{Swift}/BAT light curve of SAX1753 and found that the source underwent a faint outburst in 2012 June/July that was not reported (Fig. \ref{SAX1753_BAT}). Our \emph{Gemini} observations took place close to the end of the decay of this outburst, and it is likely that there was still accretion activity at this time, which would explain why we measured a magnitude above the UKIDSS limit. 

\begin{figure}
	\centering
	\includegraphics[width=0.5\textwidth]{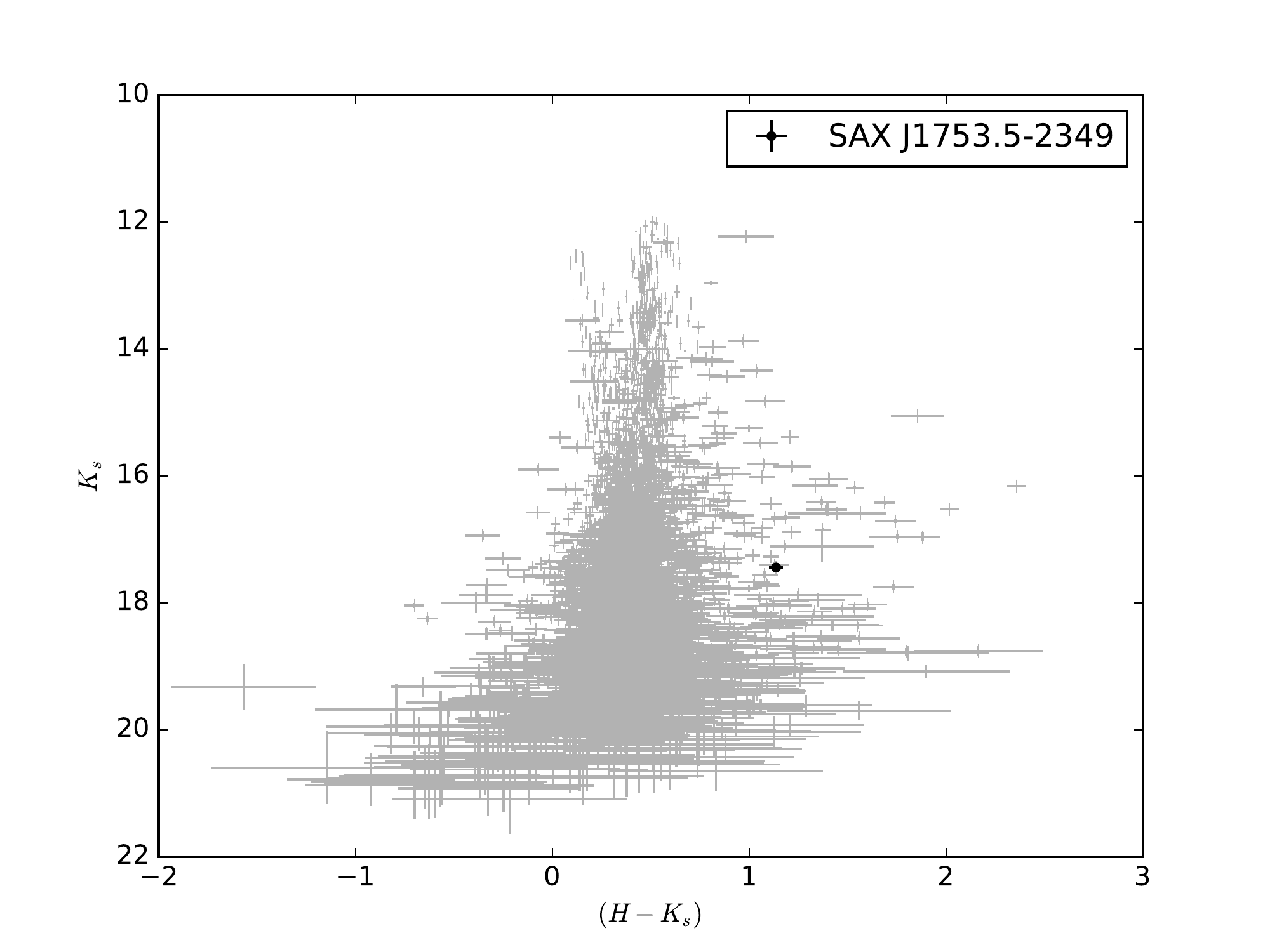}
	\caption{Colour-magnitude diagram of the SAX1753 field. The black point marks the NIRI observation of the NIR counterpart to SAX1753.}
	\label{SAX1753_colmag}
\end{figure}

\indent We present a colour-magnitude diagram of the SAX1753 field in Fig. \ref{SAX1753_colmag}. Due to the long exposure times, field stars with $K_s<12$ saturated the CCD and hence have been excluded. The NIR counterpart to SAX1753 appears to be significantly redder in relation to the field population indicating a possibility of absorption intrinsic to the system. To investigate this we calculated the NIR colour excess $E(H-K) = (H-K)_{\mathrm{obs}} - (H-K)_{\mathrm{int}}$ using the derived relationships between the total extinction in the $V$-band ($A_V$) and the $H$ and $K$-bands \citep[$A_H=0.176A_V,A_K=0.108A_V$;][]{Cardelli-1989,Cox-2000}, as well as the relation $N_H/A_V=(2.21\pm0.09) \times10^{21}$ cm$^{-2}$ mag$^{-1}$ \citep{Guver-2009}, where $N_H$ is the observed hydrogen column density. Utilizing $N_H=(1.9\pm0.4) \times10^{22}$ cm$^{-2}$, measured from fitting a thermal Comptonization model to the X-ray spectrum \citep{DelSanto-2010}, we determine a colour excess $E(H-K)=0.58\pm0.12$. To determine the intrinsic reddening we must convert the observed $(H-K_s)$ to $(H-K)$ using the known filter transformations\footnote{http://www.astro.caltech.edu/~jmc/2mass/v3/transformations/} between 2MASS/\emph{Gemini} and the \citet{Bessell-1988} photometric system. We determine $(H-K)_{\mathrm{obs}}=1.13\pm0.05$ mag and therefore an intrinsic colour index of  $(H-K)_{\mathrm{int}}=0.55\pm0.13$ mag.\\
\indent To determine whether this intrinsic reddening is typical we match the coordinates of all field sources with $(H-K_s)_{\mathrm{obs}}>1$ against the field image. We find an isotropic distribution of such sources in the field, indicating that SAX1753 is not located in a region of high absorption (e.g. behind a dust cloud) and it is possible that there is absorption local to the system itself. It is important to note that the calculation is dependent on the determination of $N_H$, which is highly dependent on the model used to define the X-ray continuum, as well as the absorption model utilized during spectral fitting. It is also possible that, as SAX1753 was potentially still showing signs of accretion at the time of the NIRI observations, the intrinsically red spectrum could be due to an outflow in the form of an accretion disc wind, or to jets - which are known to exhibit optically thin (negative slope) spectra in the infrared \citep[see e.g.][]{Diaz-Trigo-2017}.\\
\indent If we assume that the secondary star fills its Roche lobe, then we can use the NIR magnitude from the 2010 outburst to constrain the orbital period of the system using the relationship between $K$-band magnitude, $L_X$ and $P_{{\rm orb}}$ derived by \citet{Revnivtsev-2012}. Assuming a distance of 8 kpc, a luminosity of $L_X\approx0.02L_{{\rm Edd}}$ (for a $1.4M_{\odot}$ NS) at the peak of the outburst \citep{DelSanto-2010} and $K=15.63$ \citep{Torres-2010} we estimate $P_{{\rm orb}}\approx15$h with a typical propagated uncertainty of $\pm4$h. There are a number of assumptions in this calculation, for example we assume here that all of the NIR emission is due to reprocessed X-rays and that at the peak of the outburst there is no contribution from a jet or the companion (which may not be a reasonable assumption as discussed above). However, this estimate of $P_{{\rm orb}}$ argues that the system is not ultracompact, though the only way to truly determine $P_{{\rm orb}}$ is through quiescent time-series photometry or spectroscopy.\\

\subsection{SAX J1806.5--2215}

\begin{figure*}
	\centering
	\includegraphics[width=0.75\textwidth]{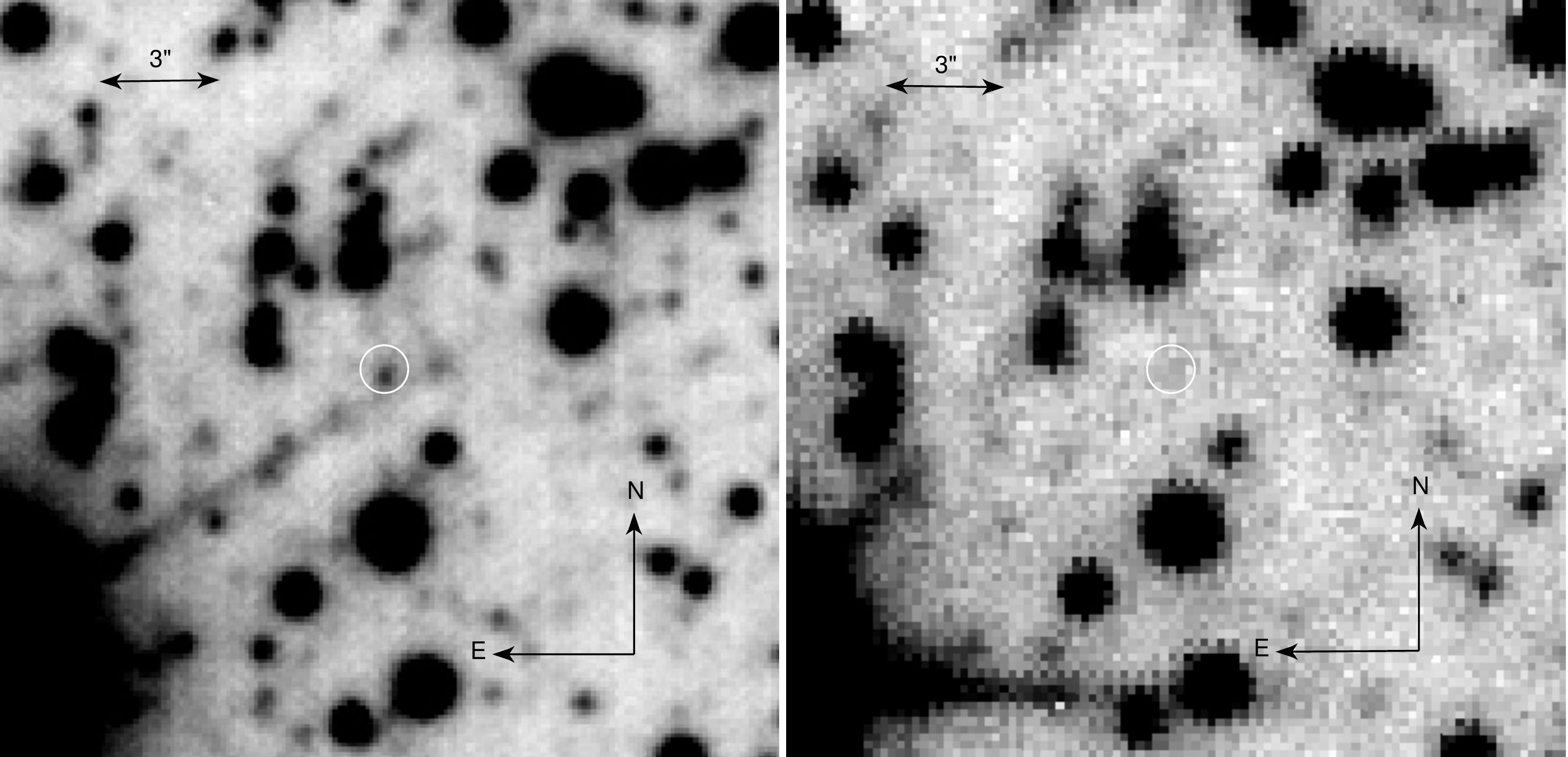}
	\caption{\emph{left:} $K_s$-band NIRI image of SAX1806. \emph{Right:} $K$-band image of the same field, observed with \emph{UKIRT} on 2006 July 23 as part of the UKIDSS. In both panels, the white circle represents the 0.6$''$ error circle of the \emph{Chandra} X-ray position of the source \citep{Chakrabarty-2011}.}
	\label{SAX1806_image}
\end{figure*}

\begin{figure}
	\centering
	\includegraphics[width=0.48\textwidth]{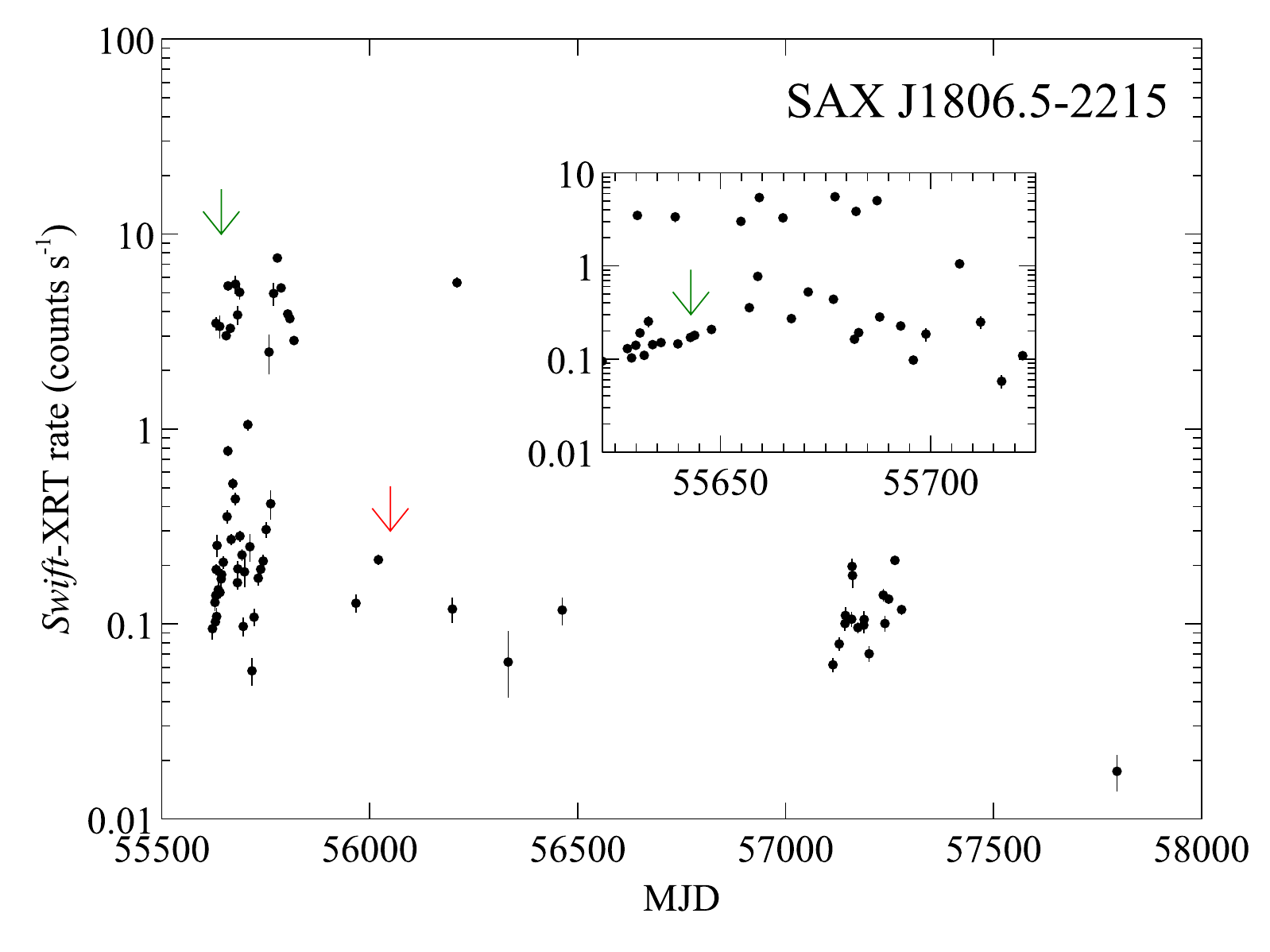}
	\caption{Long term \emph{Swift}/XRT light curve of SAX1806, the green arrow represents the time of the observation presented by \citet{Kaur-2017}, the red arrow indicates the time of the \emph{Gemini}/NIRI observation. A zoomed in portion of the light curve is shown inset.}
	\label{SAX1806_Swift}
\end{figure}

Fig. \ref{SAX1806_image} shows the $K_s$-band field of SAX1806 as observed by \emph{Gemini}/NIRI. We detect the source at a position of RA, Dec = $18^{\mathrm{h}}06^{\mathrm{m}}32^{\mathrm{s}}.18$, $-22^{\circ}14'17''.36$, within the \emph{Chandra} error circle \citep{Chakrabarty-2011}. We measure magnitudes of the NIR counterpart of $H=17.94\pm0.06$ and $K_s=17.22\pm0.02$. The measured $K_s$ magnitude is consistent with 2011 $K$-band observations of the source \citep{Kaur-2017}. Fig. \ref{SAX1806_Swift} shows the long term \emph{Swift}/XRT light curve. Though it initially seems as if the X-ray count rate had decayed significantly between the observations of \citet{Kaur-2017} and those of this work, the inset shows that the two NIR observations occurred at similar X-ray fluxes, consistent with the near identical NIR magnitudes observed at the two epochs. SAX1806 is not detected in UKIDSS, which observed the field in July 2006 when SAX1806 was in quiescence \citep{Campana-2009}, placing a $5\sigma$ upper limit on the quiescent magnitude of $K>18.05$, $H>19.00$.\\
\indent It is important to note that the NIR counterpart to SAX1806 is located close to a diffraction spike of a nearby bright star (2MASS 18063303-2214249), which may have an effect on the measured magnitude of the target. However, an investigation of the PSF subtracted image revealed no obvious residuals at the location of the source, and the diffraction spike was still present, indicative of a clean subtraction. In addition, we measured the magnitudes of two known sources present in the diffraction spikes (UGPS J180633.87--221431.0 and UGPS J180633.67--221429.3) and found them to be consistent with measured UKIDSS magnitudes. The same stars are not located in the diffraction spikes in the archival UKIDSS images. We can therefore assume that our PSF photometry is unaffected by the presence of the diffraction spike of the bright star.

\begin{figure}
	\centering
	\includegraphics[width=0.5\textwidth]{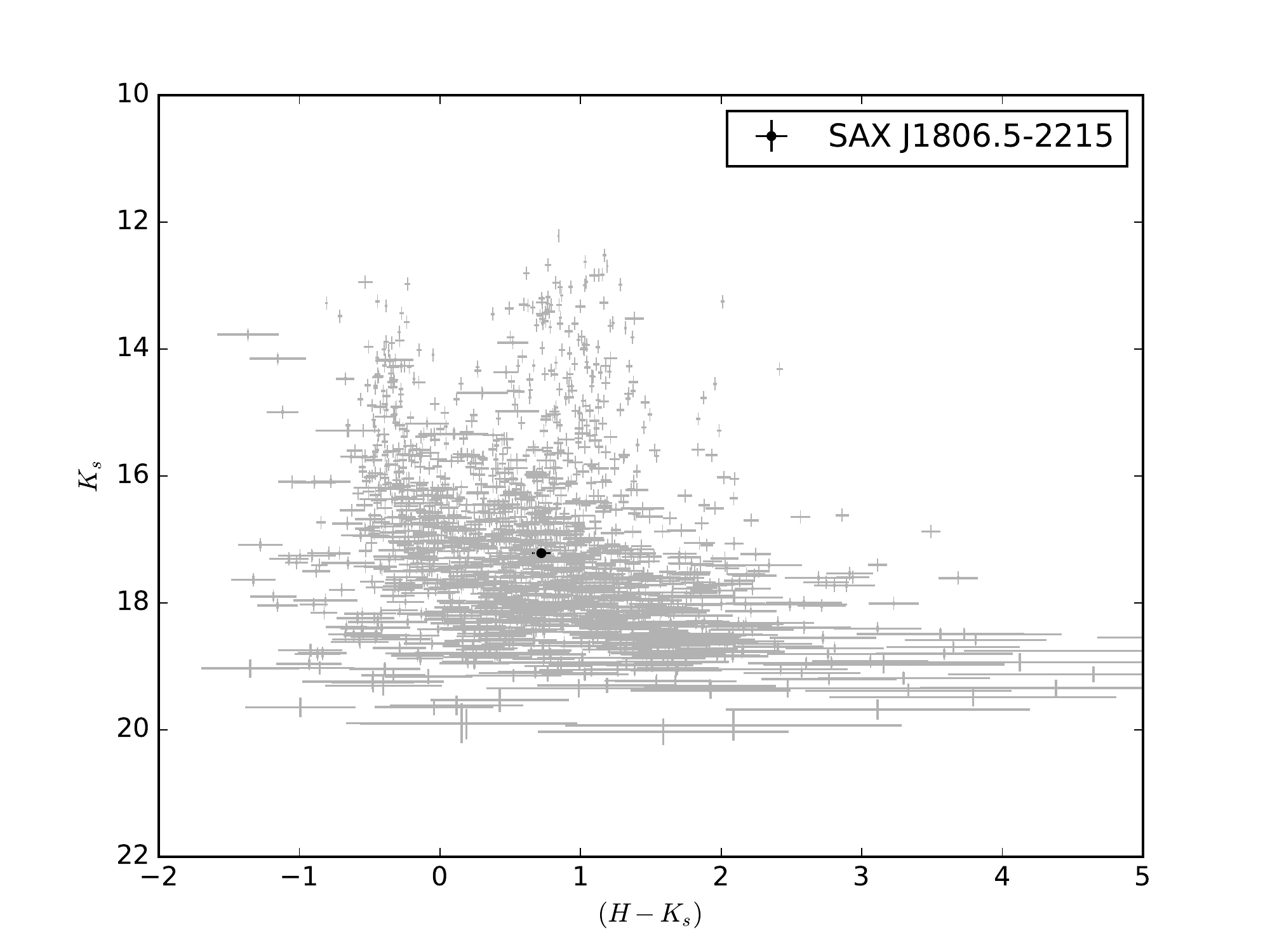}
	\caption{Colour-magnitude diagram of the SAX1806 field. The black point marks the NIRI observation of the NIR counterpart to SAX1806.}
	\label{SAX1806_colmag}
\end{figure}

\begin{figure}
	\centering
	\includegraphics[width=0.5\textwidth]{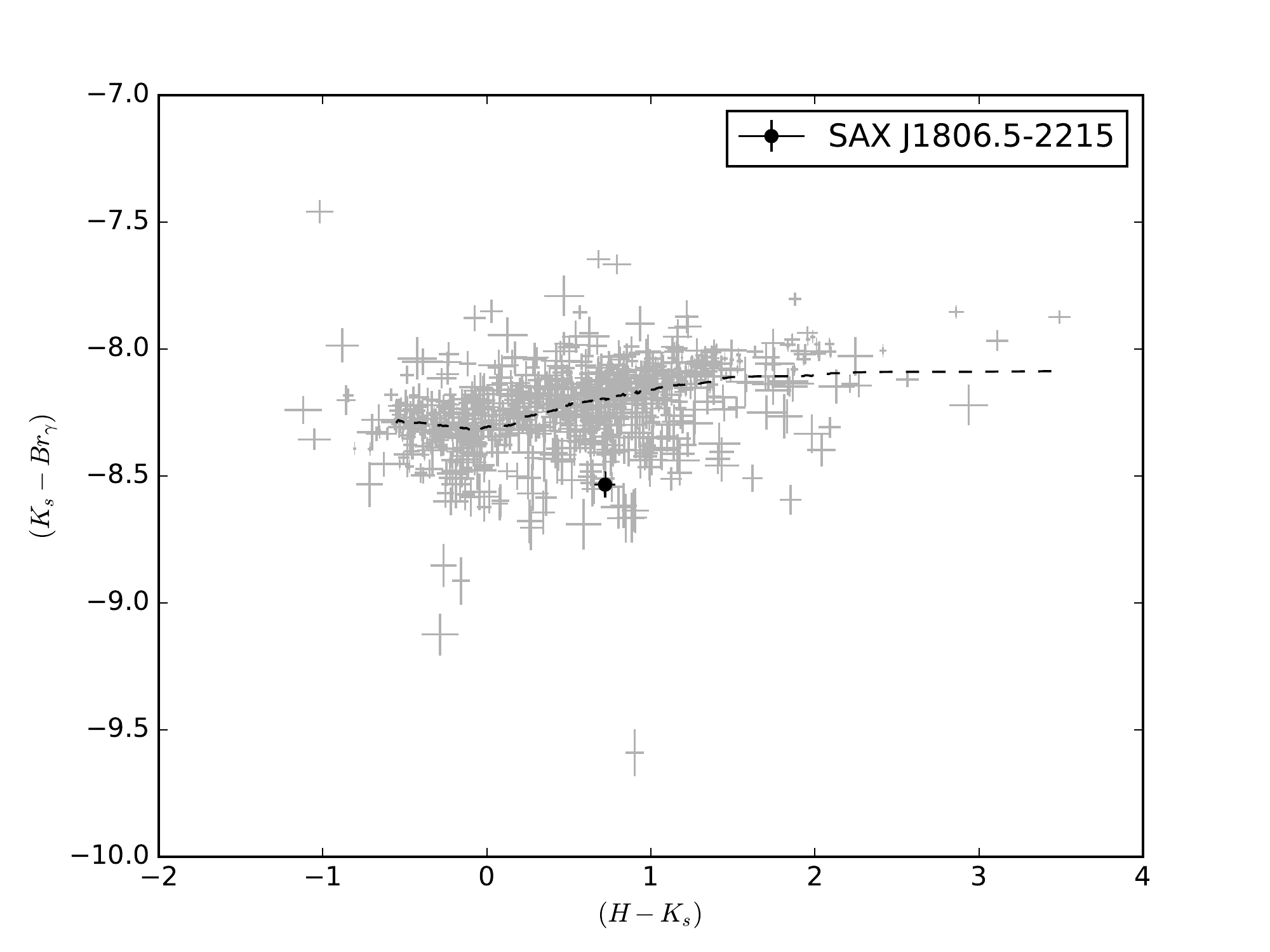}
	\caption{($H-K_s$) vs. ($K_s-$Br$_{\gamma}$) colour-colour diagram of the SAX1806 field. The black point marks the NIRI observation of the NIR counterpart to SAX1806. The dashed line represents a running median and defines the locus at which stars have no excess Br$_{\gamma}$ emission. }
	\label{SAX1806_colcol}
\end{figure}

\indent As with SAX1753, we present a colour-magnitude diagram of the source field of SAX1806 in Fig. \ref{SAX1806_colmag}. The source reddening is consistent with the majority of the field population. We also obtained observations of the source field in the Br$_{\gamma}$ narrow-band filter, which allows us to study the hydrogen emission in the system.\\
\indent It is possible to identify stars with strong line emission/absorption by using a broadband photometric filter as an indicator of the continuum near the location of the spectral line \citep[see e.g.][for methods of finding H$_{\alpha}$ emitting stars using photometry]{Robertson-1989,Panagia-2000,Drew-2005}. In the case of SAX1806, we use the $K_s$-band magnitude as an indicator of the continuum near the Br$_{\gamma}$ line. Therefore, a large ($K_s-$Br$_{\gamma}$) colour would provide evidence of excess hydrogen emission in the system.\\
\indent To construct the ($H-K_s$) vs. ($K_s-$Br$_{\gamma}$) colour-colour diagram in Fig. \ref{SAX1806_colcol} we utilised the instrumental magnitudes in the Br$_{\gamma}$ band, as we cannot convert Br$_{\gamma}$ instrumental magnitudes on to the standard photometric system due to the absence of photometric standard stars in this band. We also excluded saturated stars and stars close to the edge of the CCD. The dashed line in Fig. \ref{SAX1806_colcol} is the running median ($K_s-$Br$_{\gamma}$) colour, and represents the locus at which stars have no excess Br$_{\gamma}$ emission. Any sources lying significantly above/below this line can be assumed to exhibit a strong Br$_{\gamma}$ feature in emission/absorption.

The SAX1806 ($K_s-$Br$_{\gamma}$) colour places it well below the locus of the continuum at the same ($H-K_s$), showing a $|\Delta$Br$_{\gamma}|$ $>5$ times larger than the photometric uncertainty on the ($K_s-$Br$_{\gamma}$) colour of SAX1806. This is evidence that the source is exhibiting significant Br$_{\gamma}$ absorption.\\ 
\indent We can attempt to place an estimate on the equivalent width of the Br$_{\gamma}$ line, EW(Br$_{\gamma}$) utilizing the methods of \citet{DeMarchi-2010,Beccari-2014}, who derived the equivalent width of H$_{\alpha}$ to be related to the rectangular width, RW, of the H$_{\alpha}$ filter as ${\rm EW(H}_{\alpha}{\rm )=RW}\times [1-10^{-0.4\times\Delta{\rm H}_{\alpha}}]$, where the ${\rm H}_{\alpha}$ excess emission is the distance $\Delta{\rm H}_{\alpha}$ from the median. Echoing this, we estimate ${\rm EW(Br}_{\gamma}{\rm)} \sim-120$\AA, where a negative EW in this case represents absorption (contrary to standard notation). This is an disturbingly large value - an order of magnitude stronger than is typically seen in LMXBs \citep[see e.g.][]{Rahoui-2014}. \\
\indent To investigate this we create two new co-added images, one from the first 5 individual frames of the observing run and one from the second 5 frames in order to determine if the large implied EW is caused by large amplitude variability of the Br$_{\gamma}$ instrumental magnitude. We find that the source is not detected (to $5\sigma$) in either of the new images, indicating that the source truly is likely to be exhibiting Br$_{\gamma}$ at a lower flux than is typical of the field population. Though we cannot truly determine the properties of the Br$_{\gamma}$ line without NIR spectroscopy, this does provide evidence that there is hydrogen present in the SAX1806 system, and therefore it is likely that it is not an ultracompact binary.\\
\indent We can examine this further by applying the \citet{Revnivtsev-2012} scaling relation to the measured flux values. Assuming a distance of 8kpc and $L_X\sim0.01L_{{\rm Edd}}$ \citep{Cornelisse-2002} we estimate $P_{{\rm orb}}\approx4\pm1$h. We make the same assumptions here as with the $P_{{\rm orb}}$ estimate for SAX1753. This provides further evidence that SAX1806 is not an ultracompact system, as discussed above.


\subsection{AX J1754.2--2754}
\label{AX1754_res}

\begin{figure*}
	\centering
	\includegraphics[width=0.75\textwidth]{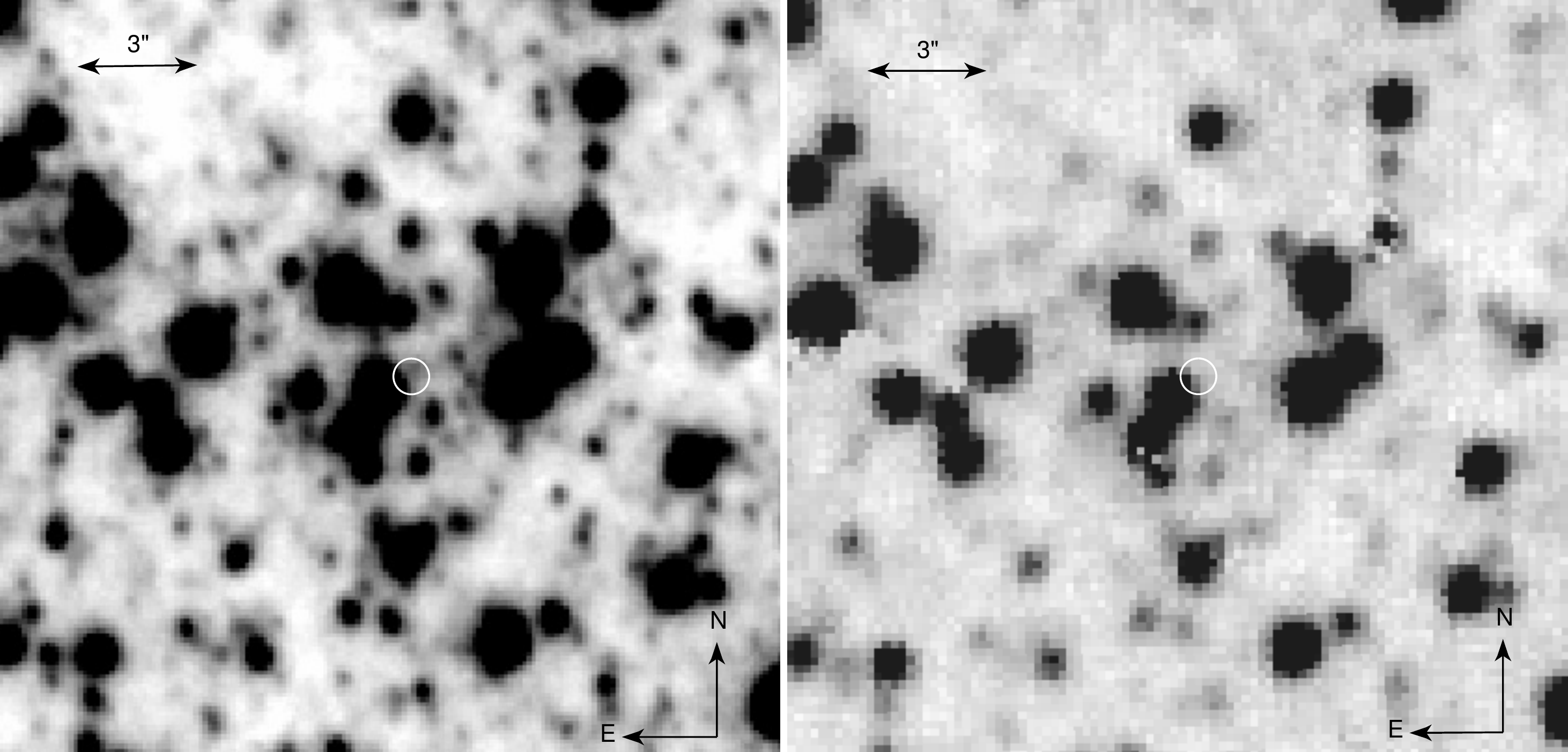}
	\caption{\emph{Left:} $K_s$-band NIRI image of AX1754 \emph{Right:} $K$-band image of the same field, observed with \emph{UKIRT} on 2006 July 23 as part of the UKIDSS. In both panels, the white circle represents the 0.45$''$ error circle of the \emph{Chandra} X-ray position of the source \citep{Bassa-2008}.}
	\label{AX1754_image}
\end{figure*}

\begin{figure}
	\centering
	\includegraphics[width=0.48\textwidth]{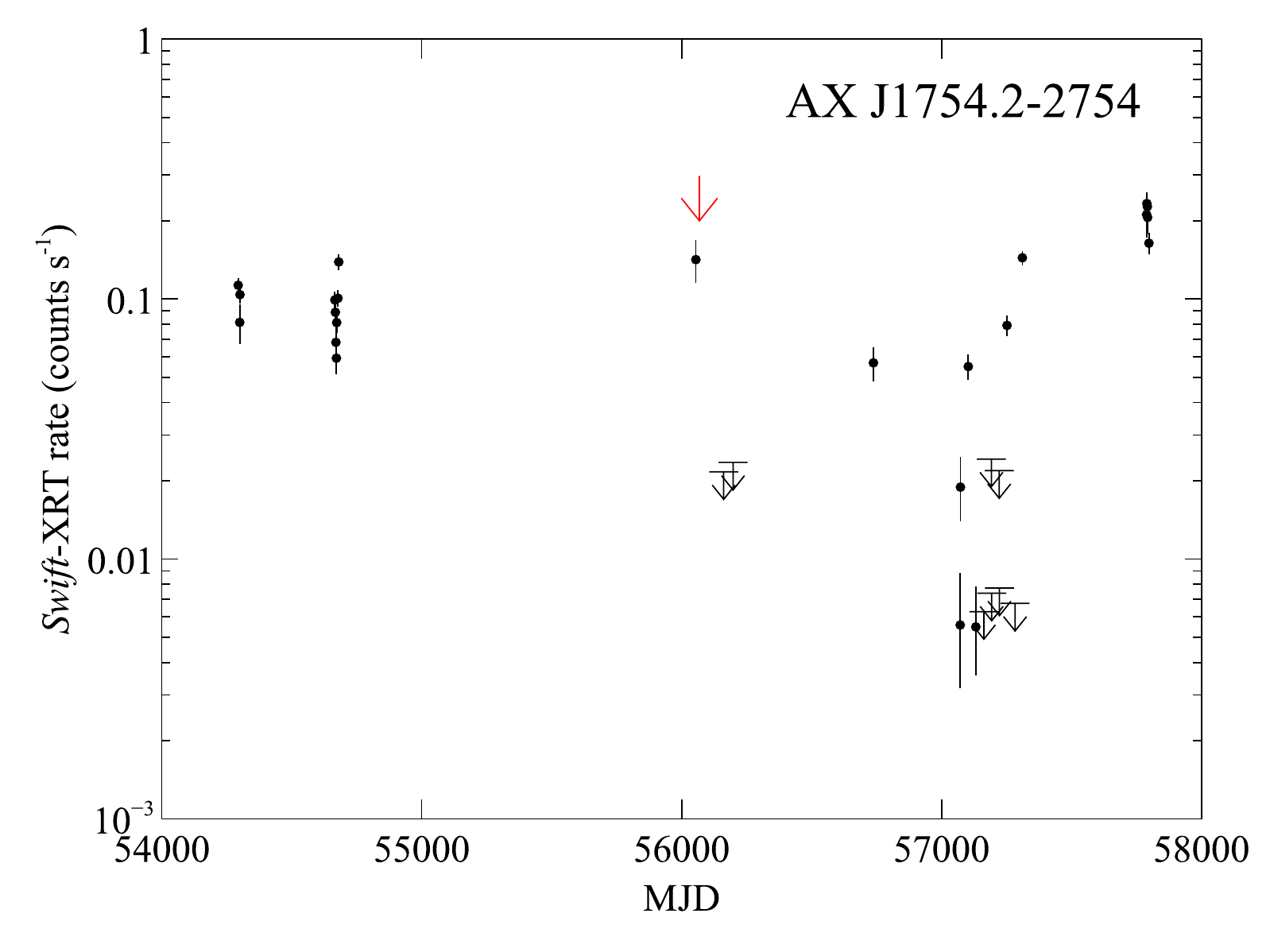}
	\caption{Long term \emph{Swift}/XRT light curve of AX1754, the red arrow indicates the time of the \emph{Gemini}/NIRI observation. Black, capped arrows represent $3\sigma$ upper limits on the XRT count rate.}
	\label{AX1754_Swift}
\end{figure}

The AX1754 field in the $K_s$-band is presented in Fig. \ref{AX1754_image}. After using the PSF fit to subtract the nearby bright star (UGPS J175414.57--275436.0) we find a source at a level of $\sim5\sigma$ above the background at a position of RA, Dec $=17^{{\rm h}}54^{{\rm m}}14^{{\rm s}}.47, -27^{\circ}54'35''.34$, well inside the \emph{Chandra} error circle \citep{Bassa-2008}. We therefore conclude that this source is the NIR counterpart of AX1754, and the measured magnitude of $K_s=18.12\pm0.15$ represents the first optical/NIR detection of this source. We do not detect a counterpart in the $H$-band (to a depth of $H=20$) or Br$_{\gamma}$-band observations.\\
\indent There is no evidence of a NIR counterpart in the archival UKIDSS image (Fig. \ref{AX1754_image}), which has a limiting magnitude of $K=18.05$ \citep{Lucas-2008}. The source was X-ray active at the time of the NIRI observations (Fig. \ref{AX1754_Swift}), so the quiescent magnitude of this source remains unknown.\\
\indent \citet{Zolotukhin-2015} provided a constraint of $P_{{\rm orb}}<9$h using the \citet{Revnivtsev-2012} relations, based on an upper limit to the NIR brightness, as the source has not been detected in archival images. We can provide a more accurate estimate using the measured $K_s$-band magnitude. Adopting $L_x\sim0.0006L_{{\rm Edd}}$ for a distance of 9.2kpc \citep{Chelovekov-2007b} we estimate $P_{{\rm orb}}\approx5.4\pm2.3$h. Though this is clearly too long for an ultracompact system, it is important to note that the calculation relies heavily on the distance measurement. Adopting the closest distance estimate calculated for AX1754 \citep[6.6kpc;][]{Chelovekov-2007b}, we find $P_{{\rm orb}}\approx2.6\pm1.1$h, which is indicative of a much more compact binary, though probably not ultracompact. We note that we have assumed that the measured $K_s$-band magnitude is the peak value. However, it is possible that AX1754 has been brighter in the NIR over the course of its prolonged X-ray activity, implying a larger $P_{{\rm orb}}$, hence making an ultracompact scenario even less likely. \\

\section{Summary and Conclusions}
\label{Discussion}
We present here NIR observations of the counterparts of three VFXTs. Included in this analysis is the first ever NIR detection of AX1754. We also present analysis of two other sources, SAX1753 and SAX1806, which have only previously been studied in single NIR bands, preventing in depth discussion of NIR colour. \\
\indent We identified the NIR counterpart to SAX1753, consistent with the position of the source reported during its 2010 outburst \citep{Torres-2010}. We find that the flux of the source has decayed by $\sim2$ magnitudes in the $K_s$-band, consistent with an apparent return to quiescence in X-rays. However, we note in Section \ref{res:1753} that SAX1753 was likely not in quiescence at the time of the NIRI observations, as a source with a quiescent magnitude of $K_s=17.44\pm0.02$ is detectable in UKIDSS archival images. The \emph{Swift}/BAT light curves in Fig. \ref{SAX1753_BAT} show evidence for a faint outburst close to the time of our NIRI observations, which suggests that the source was still exhibiting accretion activity during this time, and hence not in true quiescence.\\
\indent SAX1753 appears to exhibit significant reddening in relation to the field population (Fig. \ref{SAX1753_colmag}), with an intrinsic colour index $(H-K)_{\mathrm{int}}=0.55\pm0.13$ mag, which is larger than typical values for main sequence companions \citep{Cox-2000}. It must be noted that this value is highly dependent on the derived $N_H$, which itself is dependent on the model chosen to constrain the X-ray spectrum. We therefore cannot definitively state that there is absorption intrinsic to the system. However, it is clear from Fig. \ref{SAX1753_colmag} that SAX1753 is redder than the majority of other stars in the field, independent of the modelled $N_H$. Another possible origin for the unusually red colour of SAX1753 is the presence of a jet. It has been shown in a number of NS LMXBs that jets manifest in the NIR spectrum as a power law with a negative slope \citep{Migliari-2010,Baglio-2016,Diaz-Trigo-2017} and it is possible that we are seeing a similar process in SAX1753, as the source appeared to still be accreting at the time of our observations. However, we cannot constrain any properties with just two NIR data points, and SAX1753 has never exhibited any signs of a radio jet, so it is unclear if this interpretation is correct.\\ 
\indent We also obtained the first multi-band NIR photometry of the counterpart to SAX1806, which had previously only been observed in the $K$-band \citep{Kaur-2017}. We find that the source exhibits a low $K_s-{\rm Br}_{\gamma}$ colour in relation to the rest of the field population. This suggests the presence of a ${\rm Br}_{\gamma}$ absorption feature in the NIR spectrum of SAX1806 and, though the derived EW of $\sim120$ seems rather large, is indicative of hydrogen present in the system. This, coupled with an estimate of $P_{{\rm orb}}\approx4\pm1$h from the \citet{Revnivtsev-2012} relations provides significant evidence against an ultracompact nature. We require NIR spectroscopy to confirm the absorption in the Br$_{\gamma}$-band, and SAX1806 is bright and isolated enough to be targeted for spectroscopy with the current generation of 8m class telescopes.\\
\indent We present the first ever NIR detection of AX1754, a NS system that is actively accreting most of the time and exhibits intermediate length X-ray bursts. The $K_s$-band magnitude of $18.12\pm0.15$, combined with the extremely low X-ray luminosity, suggests a short period system ($P_{{\rm orb}}\approx5.4\pm2.3$h), though not ultracompact in nature. The source is in close proximity to a bright star, which makes NIR photometric and spectroscopic follow up extremely difficult, and hence determining the true nature of the system becomes problematic.\\
\indent We have shown in this work that the three faint NS systems presented here are likely normal LMXBs. They do not show any evidence of ultracompact behaviour, either from estimates of $P_{{\rm orb}}$ or through inferring the presence of hydrogen. In addition, several other VFXTs are found to exhibit properties typical of regular LMXBs rather than ultracompact binaries \citep[see e.g.][]{Heinke-2009,Degenaar-2010a}. However, this does not rule out the ultracompact scenario as a way of explaining the faint outbursts of a subclass of VFXTs, but rather suggests that there are multiple classes of sources accreting in a faint regime. It is worth noting here that the best studied sample of VFXTs is those that show long ($\gtrsim1$ yr) outbursts (the so-called `quasi-persistent' sources - two of which, SAX1806 and AX1754 are presented in this work), and the accretion regimes may be different among the shorter duration transients. We require dedicated photometric and spectroscopic observing campaigns to fully determine the nature of accretion in VFXTs, but this remains difficult at such low optical/NIR fluxes, and we will likely require the next generation of large telescopes to achieve this goal for a large number of sources.

\section*{Acknowledgements}
The authors thank the anonymous referee for helpful suggestions and comments which helped to improve the manuscript. COH is supported by an NSERC Discovery Grant and a Discovery Accelerator Supplement. ND is supported by a Vidi grant from the Netherlands Organization for Scientific Research (NWO). RW is supported by a NWO Top Grant, module 1. LMF acknowledges an NSERC USRA. Based on observations obtained at the \emph{Gemini} Observatory (Program IDs GN-2012A-Q-72 and GN-2012A-Q-113; acquired through the \emph{Gemini} Observatory Archive and processed using the \emph{Gemini} IRAF package), which is operated by the Association of Universities for Research in Astronomy, Inc., under a cooperative agreement with the NSF on behalf of the \emph{Gemini} partnership: the National Science Foundation (United States), the National Research Council (Canada), CONICYT (Chile), Ministerio de Ciencia, Tecnolog\'{i}a e Innovaci\'{o}n Productiva (Argentina), and Minist\'{e}rio da Ci\^{e}ncia, Tecnologia e Inova\c{c}\~{a}o (Brazil). This work made use of data supplied by the UK Swift Science Data Centre at the University of Leicester.



\bibliographystyle{mnras}
\bibliography{references.NIRILMXBS_FINAL.bib} 

\begin{thebibliography}{}
\makeatletter
\relax
\def\mn@urlcharsother{\let\do\@makeother \do\$\do\&\do\#\do\^\do\_\do\%\do\~}
\def\mn@doi{\begingroup\mn@urlcharsother \@ifnextchar [ {\mn@doi@}
  {\mn@doi@[]}}
\def\mn@doi@[#1]#2{\def\@tempa{#1}\ifx\@tempa\@empty \href
  {http://dx.doi.org/#2} {doi:#2}\else \href {http://dx.doi.org/#2} {#1}\fi
  \endgroup}
\def\mn@eprint#1#2{\mn@eprint@#1:#2::\@nil}
\def\mn@eprint@arXiv#1{\href {http://arxiv.org/abs/#1} {{\tt arXiv:#1}}}
\def\mn@eprint@dblp#1{\href {http://dblp.uni-trier.de/rec/bibtex/#1.xml}
  {dblp:#1}}
\def\mn@eprint@#1:#2:#3:#4\@nil{\def\@tempa {#1}\def\@tempb {#2}\def\@tempc
  {#3}\ifx \@tempc \@empty \let \@tempc \@tempb \let \@tempb \@tempa \fi \ifx
  \@tempb \@empty \def\@tempb {arXiv}\fi \@ifundefined
  {mn@eprint@\@tempb}{\@tempb:\@tempc}{\expandafter \expandafter \csname
  mn@eprint@\@tempb\endcsname \expandafter{\@tempc}}}

\bibitem[\protect\citeauthoryear{{Altamirano}, {Kaur}, {Degenaar}, {Wijnands},
  {Yang}, {Armas-Padilla}, {Strohmayer}  \& {Markwardt}}{{Altamirano}
  et~al.}{2011}]{Altamirano-2011}
{Altamirano} D.,  {Kaur} R.,  {Degenaar} N.,  {Wijnands} R.,  {Yang} Y.,
  {Armas-Padilla} M.,  {Strohmayer} T.,   {Markwardt} C.,  2011, The
  Astronomer's Telegram, \href
  {http://adsabs.harvard.edu/abs/2011ATel.3193....1A} {3193}

\bibitem[\protect\citeauthoryear{{Armas Padilla}, {Degenaar}  \&
  {Wijnands}}{{Armas Padilla} et~al.}{2013}]{Armas-Padilla-2013}
{Armas Padilla} M.,  {Degenaar} N.,   {Wijnands} R.,  2013, \mn@doi [\mnras]
  {10.1093/mnras/stt1114}, \href
  {http://adsabs.harvard.edu/abs/2013MNRAS.434.1586A} {434, 1586}

\bibitem[\protect\citeauthoryear{{Arnason}, {Sivakoff}, {Heinke}, {Cohn}  \&
  {Lugger}}{{Arnason} et~al.}{2015}]{Arnason-2015}
{Arnason} R.~M.,  {Sivakoff} G.~R.,  {Heinke} C.~O.,  {Cohn} H.~N.,   {Lugger}
  P.~M.,  2015, \mn@doi [\apj] {10.1088/0004-637X/807/1/52}, \href
  {http://adsabs.harvard.edu/abs/2015ApJ...807...52A} {807, 52}

\bibitem[\protect\citeauthoryear{{Baglio}, {D'Avanzo}, {Campana}, {Goldoni},
  {Masetti}, {Mu{\~n}oz-Darias}, {Pati{\~n}o-{\'A}lvarez}  \&
  {Chavushyan}}{{Baglio} et~al.}{2016}]{Baglio-2016}
{Baglio} M.~C.,  {D'Avanzo} P.,  {Campana} S.,  {Goldoni} P.,  {Masetti} N.,
  {Mu{\~n}oz-Darias} T.,  {Pati{\~n}o-{\'A}lvarez} V.,   {Chavushyan} V.,
  2016, \mn@doi [\aap] {10.1051/0004-6361/201527147}, \href
  {http://adsabs.harvard.edu/abs/2016A%26A...587A.102B} {587, A102}

\bibitem[\protect\citeauthoryear{{Bassa} et~al.,}{{Bassa}
  et~al.}{2008}]{Bassa-2008}
{Bassa} C.,  et~al., 2008, The Astronomer's Telegram, \href
  {http://adsabs.harvard.edu/abs/2008ATel.1575....1B} {1575}

\bibitem[\protect\citeauthoryear{{Beccari}, {De Marchi}, {Panagia}  \&
  {Pasquini}}{{Beccari} et~al.}{2014}]{Beccari-2014}
{Beccari} G.,  {De Marchi} G.,  {Panagia} N.,   {Pasquini} L.,  2014, \mn@doi
  [\mnras] {10.1093/mnras/stt2074}, \href
  {http://adsabs.harvard.edu/abs/2014MNRAS.437.2621B} {437, 2621}

\bibitem[\protect\citeauthoryear{{Bertin}}{{Bertin}}{2006}]{Bertin-2006}
{Bertin} E.,  2006, in {Gabriel} C.,  {Arviset} C.,  {Ponz} D.,   {Enrique} S.,
   eds,  Astronomical Society of the Pacific Conference Series Vol. 351,
  Astronomical Data Analysis Software and Systems XV. p.~112

\bibitem[\protect\citeauthoryear{{Bertin} \& {Arnouts}}{{Bertin} \&
  {Arnouts}}{1996}]{Bertin-1996}
{Bertin} E.,  {Arnouts} S.,  1996, \mn@doi [\aaps] {10.1051/aas:1996164}, \href
  {http://adsabs.harvard.edu/abs/1996A%26AS..117..393B} {117, 393}

\bibitem[\protect\citeauthoryear{{Bertin}, {Mellier}, {Radovich}, {Missonnier},
  {Didelon}  \& {Morin}}{{Bertin} et~al.}{2002}]{Bertin-2002}
{Bertin} E.,  {Mellier} Y.,  {Radovich} M.,  {Missonnier} G.,  {Didelon} P.,
  {Morin} B.,  2002, in {Bohlender} D.~A.,  {Durand} D.,   {Handley} T.~H.,
  eds,  Astronomical Society of the Pacific Conference Series Vol. 281,
  Astronomical Data Analysis Software and Systems XI. p.~228

\bibitem[\protect\citeauthoryear{{Bessell} \& {Brett}}{{Bessell} \&
  {Brett}}{1988}]{Bessell-1988}
{Bessell} M.~S.,  {Brett} J.~M.,  1988, \mn@doi [\pasp] {10.1086/132281}, \href
  {http://adsabs.harvard.edu/abs/1988PASP..100.1134B} {100, 1134}

\bibitem[\protect\citeauthoryear{{Burrows} et~al.,}{{Burrows}
  et~al.}{2005}]{Burrows-2005}
{Burrows} D.~N.,  et~al., 2005, \mn@doi [\ssr] {10.1007/s11214-005-5097-2},
  \href {http://adsabs.harvard.edu/abs/2005SSRv..120..165B} {120, 165}

\bibitem[\protect\citeauthoryear{{Cadolle Bel}, {Kuulkers}, {Chenevez},
  {Beckmann}  \& {Soldi}}{{Cadolle Bel} et~al.}{2008}]{CadolleBel-2008}
{Cadolle Bel} M.,  {Kuulkers} E.,  {Chenevez} J.,  {Beckmann} V.,   {Soldi} S.,
   2008, The Astronomer's Telegram, \href
  {http://adsabs.harvard.edu/abs/2008ATel.1810....1C} {1810}

\bibitem[\protect\citeauthoryear{{Campana}}{{Campana}}{2009}]{Campana-2009}
{Campana} S.,  2009, \mn@doi [\apj] {10.1088/0004-637X/699/2/1144}, \href
  {http://adsabs.harvard.edu/abs/2009ApJ...699.1144C} {699, 1144}

\bibitem[\protect\citeauthoryear{{Cardelli}, {Clayton}  \& {Mathis}}{{Cardelli}
  et~al.}{1989}]{Cardelli-1989}
{Cardelli} J.~A.,  {Clayton} G.~C.,   {Mathis} J.~S.,  1989, \mn@doi [\apj]
  {10.1086/167900}, \href {http://adsabs.harvard.edu/abs/1989ApJ...345..245C}
  {345, 245}

\bibitem[\protect\citeauthoryear{{Chakrabarty}, {Jonker}  \&
  {Markwardt}}{{Chakrabarty} et~al.}{2010}]{Chakrabarty-2010}
{Chakrabarty} D.,  {Jonker} P.~G.,   {Markwardt} C.~B.,  2010, The Astronomer's
  Telegram, \href {http://adsabs.harvard.edu/abs/2010ATel.2540....1C} {2540}

\bibitem[\protect\citeauthoryear{{Chakrabarty}, {Jonker}  \&
  {Markwardt}}{{Chakrabarty} et~al.}{2011}]{Chakrabarty-2011}
{Chakrabarty} D.,  {Jonker} P.,   {Markwardt} C.~B.,  2011, The Astronomer's
  Telegram, \href {http://adsabs.harvard.edu/abs/2011ATel.3218....1C} {3218}

\bibitem[\protect\citeauthoryear{Chelovekov \& Grebenev}{Chelovekov \&
  Grebenev}{2007a}]{Chelovekov-2007b}
Chelovekov I.~V.,  Grebenev S.~A.,  2007a, \mn@doi [Astronomy Letters]
  {10.1134/S1063773707120043}, 33, 807

\bibitem[\protect\citeauthoryear{{Chelovekov} \& {Grebenev}}{{Chelovekov} \&
  {Grebenev}}{2007b}]{Chelovekov-2007a}
{Chelovekov} I.~V.,  {Grebenev} S.~A.,  2007b, The Astronomer's Telegram, \href
  {http://adsabs.harvard.edu/abs/2007ATel.1094....1C} {1094}

\bibitem[\protect\citeauthoryear{{Chenevez} et~al.,}{{Chenevez}
  et~al.}{2010}]{Chenevez-2010}
{Chenevez} J.,  et~al., 2010, The Astronomer's Telegram, \href
  {http://adsabs.harvard.edu/abs/2010ATel.2505....1C} {2505}

\bibitem[\protect\citeauthoryear{{Chenevez} et~al.,}{{Chenevez}
  et~al.}{2017}]{Chenevez-2017}
{Chenevez} J.,  et~al., 2017, The Astronomer's Telegram, No.~10195, \href
  {http://adsabs.harvard.edu/abs/2017ATel10195....1C} {195}

\bibitem[\protect\citeauthoryear{{Cooper} \& {Narayan}}{{Cooper} \&
  {Narayan}}{2007}]{Cooper-2007}
{Cooper} R.~L.,  {Narayan} R.,  2007, \mn@doi [\apj] {10.1086/513461}, \href
  {http://adsabs.harvard.edu/abs/2007ApJ...661..468C} {661, 468}

\bibitem[\protect\citeauthoryear{{Coriat}, {Fender}  \& {Dubus}}{{Coriat}
  et~al.}{2012}]{Coriat-2012}
{Coriat} M.,  {Fender} R.~P.,   {Dubus} G.,  2012, \mn@doi [\mnras]
  {10.1111/j.1365-2966.2012.21339.x}, \href
  {http://adsabs.harvard.edu/abs/2012MNRAS.424.1991C} {424, 1991}

\bibitem[\protect\citeauthoryear{{Cornelisse} et~al.,}{{Cornelisse}
  et~al.}{2002a}]{Cornelisse-2002}
{Cornelisse} R.,  et~al., 2002a, \mn@doi [\aap] {10.1051/0004-6361:20020707},
  \href {http://adsabs.harvard.edu/abs/2002A%26A...392..885C} {392, 885}

\bibitem[\protect\citeauthoryear{{Cornelisse}, {Verbunt}, {in't Zand},
  {Kuulkers}  \& {Heise}}{{Cornelisse} et~al.}{2002b}]{Cornelisse-2002b}
{Cornelisse} R.,  {Verbunt} F.,  {in't Zand} J.~J.~M.,  {Kuulkers} E.,
  {Heise} J.,  2002b, \mn@doi [\aap] {10.1051/0004-6361:20021183}, \href
  {http://adsabs.harvard.edu/abs/2002A%26A...392..931C} {392, 931}

\bibitem[\protect\citeauthoryear{Corral-Santana, Casares, Mu{\~n}oz-Darias,
  Rodr{\'\i}guez-Gil, Shahbaz, Torres, Zurita  \& Tyndall}{Corral-Santana
  et~al.}{2013}]{Corral-Santana-2013}
Corral-Santana J.~M.,  Casares J.,  Mu{\~n}oz-Darias T.,  Rodr{\'\i}guez-Gil
  P.,  Shahbaz T.,  Torres M. A.~P.,  Zurita C.,   Tyndall A.~A.,  2013,
  \mn@doi [Science] {10.1126/science.1228222}, 339, 1048

\bibitem[\protect\citeauthoryear{{Cox}}{{Cox}}{2000}]{Cox-2000}
{Cox} A.~N.,  2000, {Allen's astrophysical quantities}

\bibitem[\protect\citeauthoryear{{D'Angelo} \& {Spruit}}{{D'Angelo} \&
  {Spruit}}{2012}]{DAngelo-2012}
{D'Angelo} C.~R.,  {Spruit} H.~C.,  2012, \mn@doi [\mnras]
  {10.1111/j.1365-2966.2011.20046.x}, \href
  {http://esoads.eso.org/abs/2012MNRAS.420..416D} {420, 416}

\bibitem[\protect\citeauthoryear{{De Marchi}, {Panagia}  \& {Romaniello}}{{De
  Marchi} et~al.}{2010}]{DeMarchi-2010}
{De Marchi} G.,  {Panagia} N.,   {Romaniello} M.,  2010, \mn@doi [\apj]
  {10.1088/0004-637X/715/1/1}, \href
  {http://adsabs.harvard.edu/abs/2010ApJ...715....1D} {715, 1}

\bibitem[\protect\citeauthoryear{{Degenaar} \& {Wijnands}}{{Degenaar} \&
  {Wijnands}}{2009}]{Degenaar-2009}
{Degenaar} N.,  {Wijnands} R.,  2009, \mn@doi [\aap]
  {10.1051/0004-6361:200810654}, \href
  {http://adsabs.harvard.edu/abs/2009A%26A...495..547D} {495, 547}

\bibitem[\protect\citeauthoryear{{Degenaar} \& {Wijnands}}{{Degenaar} \&
  {Wijnands}}{2010}]{Degenaar-2010b}
{Degenaar} N.,  {Wijnands} R.,  2010, \mn@doi [\aap]
  {10.1051/0004-6361/201015322}, \href
  {http://adsabs.harvard.edu/abs/2010A%26A...524A..69D} {524, A69}

\bibitem[\protect\citeauthoryear{{Degenaar} et~al.,}{{Degenaar}
  et~al.}{2010}]{Degenaar-2010a}
{Degenaar} N.,  et~al., 2010, \mn@doi [\mnras]
  {10.1111/j.1365-2966.2010.16388.x}, \href
  {http://adsabs.harvard.edu/abs/2010MNRAS.404.1591D} {404, 1591}

\bibitem[\protect\citeauthoryear{{Degenaar}, {Altamirano}, {Padilla}, {Kaur},
  {Wijnands}  \& {Yang}}{{Degenaar} et~al.}{2011}]{Degenaar-2011}
{Degenaar} N.,  {Altamirano} D.,  {Padilla} M.~A.,  {Kaur} R.,  {Wijnands} R.,
   {Yang} Y.~J.,  2011, The Astronomer's Telegram, \href
  {http://adsabs.harvard.edu/abs/2011ATel.3202....1D} {3202}

\bibitem[\protect\citeauthoryear{{Degenaar} et~al.,}{{Degenaar}
  et~al.}{2012}]{Degenaar-2012}
{Degenaar} N.,  et~al., 2012, \mn@doi [\aap] {10.1051/0004-6361/201118634},
  \href {http://adsabs.harvard.edu/abs/2012A%26A...540A..22D} {540, A22}

\bibitem[\protect\citeauthoryear{{Del Santo}, {Romano}  \& {Sidoli}}{{Del
  Santo} et~al.}{2009}]{DelSanto-2009}
{Del Santo} M.,  {Romano} P.,   {Sidoli} L.,  2009, The Astronomer's Telegram,
  \href {http://adsabs.harvard.edu/abs/2009ATel.1975....1D} {1975}

\bibitem[\protect\citeauthoryear{{Del Santo}, {Sidoli}, {Romano}, {Bazzano},
  {Wijnands}, {Degenaar}  \& {Mereghetti}}{{Del Santo}
  et~al.}{2010}]{DelSanto-2010}
{Del Santo} M.,  {Sidoli} L.,  {Romano} P.,  {Bazzano} A.,  {Wijnands} R.,
  {Degenaar} N.,   {Mereghetti} S.,  2010, \mn@doi [\mnras]
  {10.1111/j.1745-3933.2010.00821.x}, \href
  {http://adsabs.harvard.edu/abs/2010MNRAS.403L..89D} {403, L89}

\bibitem[\protect\citeauthoryear{{Del Santo} et~al.,}{{Del Santo}
  et~al.}{2012}]{DelSanto-2012}
{Del Santo} M.,  et~al., 2012, The Astronomer's Telegram, \href
  {http://adsabs.harvard.edu/abs/2012ATel.4017....1D} {4017}

\bibitem[\protect\citeauthoryear{{D{\'{\i}}az Trigo}, {Migliari},
  {Miller-Jones}, {Rahoui}, {Russell}  \& {Tudor}}{{D{\'{\i}}az Trigo}
  et~al.}{2017}]{Diaz-Trigo-2017}
{D{\'{\i}}az Trigo} M.,  {Migliari} S.,  {Miller-Jones} J.~C.~A.,  {Rahoui} F.,
   {Russell} D.~M.,   {Tudor} V.,  2017, \mn@doi [\aap]
  {10.1051/0004-6361/201629472}, \href
  {http://adsabs.harvard.edu/abs/2017A%26A...600A...8D} {600, A8}

\bibitem[\protect\citeauthoryear{{Drew} et~al.,}{{Drew}
  et~al.}{2005}]{Drew-2005}
{Drew} J.~E.,  et~al., 2005, \mn@doi [\mnras]
  {10.1111/j.1365-2966.2005.09330.x}, \href
  {http://adsabs.harvard.edu/abs/2005MNRAS.362..753D} {362, 753}

\bibitem[\protect\citeauthoryear{{Evans} et~al.,}{{Evans}
  et~al.}{2009}]{Evans-2009}
{Evans} P.~A.,  et~al., 2009, \mn@doi [\mnras]
  {10.1111/j.1365-2966.2009.14913.x}, \href
  {http://adsabs.harvard.edu/abs/2009MNRAS.397.1177E} {397, 1177}

\bibitem[\protect\citeauthoryear{{Greiner}, {Sala}  \& {Kruehler}}{{Greiner}
  et~al.}{2008}]{Greiner-2008}
{Greiner} J.,  {Sala} G.,   {Kruehler} T.,  2008, The Astronomer's Telegram,
  \href {http://esoads.eso.org/abs/2008ATel.1577....1G} {1577}

\bibitem[\protect\citeauthoryear{{G{\"u}ver} \& {{\"O}zel}}{{G{\"u}ver} \&
  {{\"O}zel}}{2009}]{Guver-2009}
{G{\"u}ver} T.,  {{\"O}zel} F.,  2009, \mn@doi [\mnras]
  {10.1111/j.1365-2966.2009.15598.x}, \href
  {http://adsabs.harvard.edu/abs/2009MNRAS.400.2050G} {400, 2050}

\bibitem[\protect\citeauthoryear{{Hameury} \& {Lasota}}{{Hameury} \&
  {Lasota}}{2016}]{Hameury-2016}
{Hameury} J.-M.,  {Lasota} J.-P.,  2016, \mn@doi [\aap]
  {10.1051/0004-6361/201628434}, \href
  {http://adsabs.harvard.edu/abs/2016A%26A...594A..87H} {594, A87}

\bibitem[\protect\citeauthoryear{{Heinke}, {Cohn}  \& {Lugger}}{{Heinke}
  et~al.}{2009}]{Heinke-2009}
{Heinke} C.~O.,  {Cohn} H.~N.,   {Lugger} P.~M.,  2009, \mn@doi [\apj]
  {10.1088/0004-637X/692/1/584}, \href
  {http://adsabs.harvard.edu/abs/2009ApJ...692..584H} {692, 584}

\bibitem[\protect\citeauthoryear{{Heinke}, {Bahramian}, {Degenaar}  \&
  {Wijnands}}{{Heinke} et~al.}{2015}]{Heinke-2015}
{Heinke} C.~O.,  {Bahramian} A.,  {Degenaar} N.,   {Wijnands} R.,  2015,
  \mn@doi [\mnras] {10.1093/mnras/stu2652}, \href
  {http://adsabs.harvard.edu/abs/2015MNRAS.447.3034H} {447, 3034}

\bibitem[\protect\citeauthoryear{{Hodapp} et~al.,}{{Hodapp}
  et~al.}{2003}]{Hodapp-2003}
{Hodapp} K.~W.,  et~al., 2003, \mn@doi [\pasp] {10.1086/379669}, \href
  {http://adsabs.harvard.edu/abs/2003PASP..115.1388H} {115, 1388}

\bibitem[\protect\citeauthoryear{{In't Zand}, {Heise}, {Muller}, {Bazzano},
  {Cocchi}, {Natalucci}  \& {Ubertini}}{{In't Zand}
  et~al.}{1999}]{int-Zand-1999}
{In't Zand} J.~J.~M.,  {Heise} J.,  {Muller} J.~M.,  {Bazzano} A.,  {Cocchi}
  M.,  {Natalucci} L.,   {Ubertini} P.,  1999, \mn@doi [Nuclear Physics B
  Proceedings Supplements] {10.1016/S0920-5632(98)00214-X}, \href
  {http://adsabs.harvard.edu/abs/1999NuPhS..69..228I} {69, 228}

\bibitem[\protect\citeauthoryear{{Jonker} \& {Keek}}{{Jonker} \&
  {Keek}}{2008}]{Jonker-2008}
{Jonker} P.~G.,  {Keek} L.,  2008, The Astronomer's Telegram, \href
  {http://adsabs.harvard.edu/abs/2008ATel.1643....1J} {1643}

\bibitem[\protect\citeauthoryear{{Kaur}, {Wijnands}, {Heinke}  \&
  {Degenaar}}{{Kaur} et~al.}{2012}]{Kaur-2012b}
{Kaur} R.,  {Wijnands} R.,  {Heinke} C.,   {Degenaar} N.,  2012, The
  Astronomer's Telegram, \href
  {http://adsabs.harvard.edu/abs/2012ATel.3926....1K} {3926}

\bibitem[\protect\citeauthoryear{{Kaur}, {Wijnands}, {Kamble}, {Cackett},
  {Kutulla}, {Kaplan}  \& {Degenaar}}{{Kaur} et~al.}{2017}]{Kaur-2017}
{Kaur} R.,  {Wijnands} R.,  {Kamble} A.,  {Cackett} E.~M.,  {Kutulla} R.,
  {Kaplan} D.,   {Degenaar} N.,  2017, \mn@doi [\mnras]
  {10.1093/mnras/stw2319}, \href
  {http://adsabs.harvard.edu/abs/2017MNRAS.464..170K} {464, 170}

\bibitem[\protect\citeauthoryear{{King} \& {Wijnands}}{{King} \&
  {Wijnands}}{2006}]{King-2006}
{King} A.~R.,  {Wijnands} R.,  2006, \mn@doi [\mnras]
  {10.1111/j.1745-3933.2005.00126.x}, \href
  {http://adsabs.harvard.edu/abs/2006MNRAS.366L..31K} {366, L31}

\bibitem[\protect\citeauthoryear{{Krimm} et~al.,}{{Krimm}
  et~al.}{2013}]{Krimm-2013}
{Krimm} H.~A.,  et~al., 2013, \mn@doi [\apjs] {10.1088/0067-0049/209/1/14},
  \href {http://adsabs.harvard.edu/abs/2013ApJS..209...14K} {209, 14}

\bibitem[\protect\citeauthoryear{{Kuulkers}}{{Kuulkers}}{1998}]{Kuulkers-1998}
{Kuulkers} E.,  1998, \mn@doi [\nar] {10.1016/S1387-6473(98)00019-0}, \href
  {http://adsabs.harvard.edu/abs/1998NewAR..42....1K} {42, 1}

\bibitem[\protect\citeauthoryear{{Lasota}}{{Lasota}}{2001}]{Lasota-2001}
{Lasota} J.-P.,  2001, \mn@doi [\nar] {10.1016/S1387-6473(01)00112-9}, \href
  {http://adsabs.harvard.edu/abs/2001NewAR..45..449L} {45, 449}

\bibitem[\protect\citeauthoryear{{Lucas} et~al.,}{{Lucas}
  et~al.}{2008}]{Lucas-2008}
{Lucas} P.~W.,  et~al., 2008, \mn@doi [\mnras]
  {10.1111/j.1365-2966.2008.13924.x}, \href
  {http://adsabs.harvard.edu/abs/2008MNRAS.391..136L} {391, 136}

\bibitem[\protect\citeauthoryear{{Maccarone} \& {Patruno}}{{Maccarone} \&
  {Patruno}}{2013}]{Maccarone-2013}
{Maccarone} T.~J.,  {Patruno} A.,  2013, \mn@doi [\mnras]
  {10.1093/mnras/sts113}, \href
  {http://adsabs.harvard.edu/abs/2013MNRAS.428.1335M} {428, 1335}

\bibitem[\protect\citeauthoryear{{Maccarone} et~al.,}{{Maccarone}
  et~al.}{2012}]{Maccarone-2012}
{Maccarone} T.~J.,  et~al., 2012, The Astronomer's Telegram, \href
  {http://adsabs.harvard.edu/abs/2012ATel.4109....1M} {4109}

\bibitem[\protect\citeauthoryear{{Markwardt}, {Krimm}  \& {Swank}}{{Markwardt}
  et~al.}{2008}]{Markwardt-2008}
{Markwardt} C.~B.,  {Krimm} H.~A.,   {Swank} J.~H.,  2008, The Astronomer's
  Telegram, \href {http://adsabs.harvard.edu/abs/2008ATel.1799....1M} {1799}

\bibitem[\protect\citeauthoryear{{Mauerhan}, {Muno}, {Morris}, {Bauer},
  {Nishiyama}  \& {Nagata}}{{Mauerhan} et~al.}{2009}]{Mauerhan-2009}
{Mauerhan} J.~C.,  {Muno} M.~P.,  {Morris} M.~R.,  {Bauer} F.~E.,  {Nishiyama}
  S.,   {Nagata} T.,  2009, \mn@doi [\apj] {10.1088/0004-637X/703/1/30}, \href
  {http://esoads.eso.org/abs/2009ApJ...703...30M} {703, 30}

\bibitem[\protect\citeauthoryear{{Migliari} et~al.,}{{Migliari}
  et~al.}{2010}]{Migliari-2010}
{Migliari} S.,  et~al., 2010, \mn@doi [\apj] {10.1088/0004-637X/710/1/117},
  \href {http://adsabs.harvard.edu/abs/2010ApJ...710..117M} {710, 117}

\bibitem[\protect\citeauthoryear{{Muno}, {Pfahl}, {Baganoff}, {Brandt}, {Ghez},
  {Lu}  \& {Morris}}{{Muno} et~al.}{2005a}]{Muno-2005a}
{Muno} M.~P.,  {Pfahl} E.,  {Baganoff} F.~K.,  {Brandt} W.~N.,  {Ghez} A.,
  {Lu} J.,   {Morris} M.~R.,  2005a, \mn@doi [\apjl] {10.1086/429721}, \href
  {http://adsabs.harvard.edu/abs/2005ApJ...622L.113M} {622, L113}

\bibitem[\protect\citeauthoryear{{Muno}, {Lu}, {Baganoff}, {Brandt}, {Garmire},
  {Ghez}, {Hornstein}  \& {Morris}}{{Muno} et~al.}{2005b}]{Muno-2005b}
{Muno} M.~P.,  {Lu} J.~R.,  {Baganoff} F.~K.,  {Brandt} W.~N.,  {Garmire}
  G.~P.,  {Ghez} A.~M.,  {Hornstein} S.~D.,   {Morris} M.~R.,  2005b, \mn@doi
  [\apj] {10.1086/444586}, \href
  {http://adsabs.harvard.edu/abs/2005ApJ...633..228M} {633, 228}

\bibitem[\protect\citeauthoryear{{Panagia}, {Romaniello}, {Scuderi}  \&
  {Kirshner}}{{Panagia} et~al.}{2000}]{Panagia-2000}
{Panagia} N.,  {Romaniello} M.,  {Scuderi} S.,   {Kirshner} R.~P.,  2000,
  \mn@doi [\apj] {10.1086/309212}, \href
  {http://adsabs.harvard.edu/abs/2000ApJ...539..197P} {539, 197}

\bibitem[\protect\citeauthoryear{{Peng}, {Brown}  \& {Truran}}{{Peng}
  et~al.}{2007}]{Peng-2007}
{Peng} F.,  {Brown} E.~F.,   {Truran} J.~W.,  2007, \mn@doi [\apj]
  {10.1086/509628}, \href {http://adsabs.harvard.edu/abs/2007ApJ...654.1022P}
  {654, 1022}

\bibitem[\protect\citeauthoryear{{Pfahl}, {Rappaport}  \&
  {Podsiadlowski}}{{Pfahl} et~al.}{2002}]{Pfahl-2002}
{Pfahl} E.,  {Rappaport} S.,   {Podsiadlowski} P.,  2002, \mn@doi [\apjl]
  {10.1086/341197}, \href {http://adsabs.harvard.edu/abs/2002ApJ...571L..37P}
  {571, L37}

\bibitem[\protect\citeauthoryear{{Rahoui}, {Coriat}  \& {Lee}}{{Rahoui}
  et~al.}{2014}]{Rahoui-2014}
{Rahoui} F.,  {Coriat} M.,   {Lee} J.~C.,  2014, \mn@doi [\mnras]
  {10.1093/mnras/stu977}, \href
  {http://adsabs.harvard.edu/abs/2014MNRAS.442.1610R} {442, 1610}

\bibitem[\protect\citeauthoryear{{Revnivtsev}, {Zolotukhin}  \&
  {Meshcheryakov}}{{Revnivtsev} et~al.}{2012}]{Revnivtsev-2012}
{Revnivtsev} M.~G.,  {Zolotukhin} I.~Y.,   {Meshcheryakov} A.~V.,  2012,
  \mn@doi [\mnras] {10.1111/j.1365-2966.2012.20511.x}, \href
  {http://adsabs.harvard.edu/abs/2012MNRAS.421.2846R} {421, 2846}

\bibitem[\protect\citeauthoryear{{Robertson} \& {Jordan}}{{Robertson} \&
  {Jordan}}{1989}]{Robertson-1989}
{Robertson} T.~H.,  {Jordan} T.~M.,  1989, \mn@doi [\aj] {10.1086/115219},
  \href {http://adsabs.harvard.edu/abs/1989AJ.....98.1354R} {98, 1354}

\bibitem[\protect\citeauthoryear{{Sakano}, {Koyama}, {Murakami}, {Maeda}  \&
  {Yamauchi}}{{Sakano} et~al.}{2002}]{Sakano-2002}
{Sakano} M.,  {Koyama} K.,  {Murakami} H.,  {Maeda} Y.,   {Yamauchi} S.,  2002,
  \mn@doi [\apjs] {10.1086/324020}, \href
  {http://adsabs.harvard.edu/abs/2002ApJS..138...19S} {138, 19}

\bibitem[\protect\citeauthoryear{{Sengar}, {Tauris}, {Langer}  \&
  {Istrate}}{{Sengar} et~al.}{2017}]{Sengar-2017}
{Sengar} R.,  {Tauris} T.~M.,  {Langer} N.,   {Istrate} A.~G.,  2017, preprint,
  \href {http://adsabs.harvard.edu/abs/2017arXiv170408260S} {} (\mn@eprint
  {arXiv} {1704.08260})

\bibitem[\protect\citeauthoryear{{Skrutskie} et~al.,}{{Skrutskie}
  et~al.}{2006}]{Skrutskie-2006}
{Skrutskie} M.~F.,  et~al., 2006, \mn@doi [\aj] {10.1086/498708}, \href
  {http://adsabs.harvard.edu/abs/2006AJ....131.1163S} {131, 1163}

\bibitem[\protect\citeauthoryear{{Stetson}}{{Stetson}}{1987}]{Stetson-1987}
{Stetson} P.~B.,  1987, \mn@doi [\pasp] {10.1086/131977}, \href
  {http://adsabs.harvard.edu/abs/1987PASP...99..191S} {99, 191}

\bibitem[\protect\citeauthoryear{{Tody}}{{Tody}}{1986}]{IRAF}
{Tody} D.,  1986, in {Crawford} D.~L.,  ed.,  Society of Photo-Optical
  Instrumentation Engineers (SPIE) Conference Series Vol. 627, Instrumentation
  in astronomy VI. p.~733

\bibitem[\protect\citeauthoryear{{Torres}, {Jonker}, {Steeghs}  \&
  {Chaname}}{{Torres} et~al.}{2010}]{Torres-2010}
{Torres} M.~A.~P.,  {Jonker} P.~G.,  {Steeghs} D.,   {Chaname} J.,  2010, The
  Astronomer's Telegram, \href
  {http://adsabs.harvard.edu/abs/2010ATel.2526....1T} {2526}

\bibitem[\protect\citeauthoryear{{Wijnands} et~al.,}{{Wijnands}
  et~al.}{2006}]{Wijnands-2006}
{Wijnands} R.,  et~al., 2006, \mn@doi [\aap] {10.1051/0004-6361:20054129},
  \href {http://adsabs.harvard.edu/abs/2006A%26A...449.1117W} {449, 1117}

\bibitem[\protect\citeauthoryear{{Zolotukhin} \& {Revnivtsev}}{{Zolotukhin} \&
  {Revnivtsev}}{2015}]{Zolotukhin-2015}
{Zolotukhin} I.~Y.,  {Revnivtsev} M.~G.,  2015, \mn@doi [\mnras]
  {10.1093/mnras/stu2212}, \href
  {http://adsabs.harvard.edu/abs/2015MNRAS.446.2418Z} {446, 2418}

\bibitem[\protect\citeauthoryear{{in't Zand}, {Cumming}, {van der Sluys},
  {Verbunt}  \& {Pols}}{{in't Zand} et~al.}{2005}]{int-Zand-2005}
{in't Zand} J.~J.~M.,  {Cumming} A.,  {van der Sluys} M.~V.,  {Verbunt} F.,
  {Pols} O.~R.,  2005, \mn@doi [\aap] {10.1051/0004-6361:20053002}, \href
  {http://esoads.eso.org/abs/2005A%26A...441..675I} {441, 675}

\bibitem[\protect\citeauthoryear{{in't Zand}, {Jonker}  \& {Markwardt}}{{in't
  Zand} et~al.}{2007}]{int-Zand-2007}
{in't Zand} J.~J.~M.,  {Jonker} P.~G.,   {Markwardt} C.~B.,  2007, \mn@doi
  [\aap] {10.1051/0004-6361:20066678}, \href
  {http://esoads.eso.org/abs/2007A%26A...465..953I} {465, 953}

\makeatother
\end{thebibliography}







\bsp	
\label{lastpage}
\end{document}